\documentclass[aps,prx,twocolumn,footinbib,floatfix,superscriptaddress]{revtex4-1}
\usepackage{graphicx}
\usepackage{indentfirst}
\usepackage{braket}
\usepackage{float}
\usepackage{amsmath}
\usepackage{physics}
\usepackage{CJK}
\usepackage{esint}
\usepackage{color}
\usepackage[T1]{fontenc}
\usepackage{subfigure}
\usepackage{amsfonts}
\usepackage{footmisc}
\usepackage{scrextend}
\usepackage{multirow}
\usepackage[outdir=./]{epstopdf}

\usepackage[english]{babel}
\usepackage{url}
\usepackage{comment}
\usepackage{array}
\usepackage{subfiles}
\usepackage{tikz}
\usetikzlibrary{shapes,arrows}

\usepackage[hyperfootnotes=false]{hyperref}
\usepackage{cleveref}
\definecolor{darkblue}{rgb}{0,0,0.5}
\hypersetup{
colorlinks=true,
linkcolor=black,
filecolor=blue,
citecolor=darkblue,  
urlcolor=black,
}

\newtheorem{theorem}{Theorem}

\newtheorem{corollary}[theorem]{Corollary}

\newtheorem{lemma}[theorem]{Lemma}

\newtheorem{remark}[theorem]{Remark}

\newenvironment{proof}[1][Proof]{\noindent\textbf{#1.} }{\ \rule{0.5em}{0.5em}}

\def\be{\begin{equation}}
\def\ee{\end{equation}}
\def\ba{\begin{eqnarray}}
\def\ea{\end{eqnarray}}
\def\bal{\begin{equation}\begin{aligned}}
\def\eal{\end{aligned}\end{equation}}

\usepackage{bm}

\newcommand{\calD}{{\cal D}}
\newcommand{\calE}{{\cal E}}

\newcommand{\calL}{{\cal L}}

\newcommand{\calJ}{{\cal J}}

\newcommand{\calS}{{\cal S}}

\newcommand{\1}{^{(1)}}

\newcommand{\QZ}[1]{{{\textcolor{black}{#1}}}}

\usepackage[normalem]{ulem}

\def\bp{\begin{pmatrix}}
\def\ep{\end{pmatrix}}

\begin{document}

\title{Entanglement-Assisted Absorption Spectroscopy}

\author{Haowei Shi}
\affiliation{
James C. Wyant College of Optical Sciences, University of Arizona, Tucson, Arizona 85721, USA
}
\author{Zheshen Zhang}
\affiliation{
Department of Materials Science and Engineering, University of Arizona, Tucson, Arizona 85721, USA
}
\affiliation{
James C. Wyant College of Optical Sciences, University of Arizona, Tucson, Arizona 85721, USA
}

\author{Stefano Pirandola}
\affiliation{Department of Computer Science, University of York, York YO10 5GH, UK}

\author{Quntao Zhuang}
\email{zhuangquntao@email.arizona.edu}
\affiliation{
Department of Electrical and Computer Engineering, University of Arizona, Tucson, Arizona 85721, USA
}
\affiliation{
James C. Wyant College of Optical Sciences, University of Arizona, Tucson, Arizona 85721, USA
}

\begin{abstract} 
Spectroscopy is an important tool for probing the properties of materials, chemicals and biological samples. We design a practical transmitter-receiver system that exploits entanglement to achieve a provable quantum advantage over all spectroscopic schemes based on classical sources. To probe the absorption spectra, modelled as pattern of transmissivities among different frequency modes,  we employ broad-band signal-idler pairs in two-mode squeezed vacuum states. At the receiver side, we apply photodetection after optical parametric amplification. Finally, we perform a maximal-likehihood decision test on the measurement results, achieving orders-of-magnitude-lower error probability than the optimum classical systems in various examples, including `wine-tasting' and `drug-testing' where real molecules are considered. In detecting the presence of an absorption line, our quantum scheme achieves the optimum performance allowed by quantum mechanics. The quantum advantage in our system is robust against noise and loss, which makes near-term experimental demonstration possible.
\end{abstract} 

\maketitle

Quantum sensing and metrology~\cite{pirandola2018advances,Ruo,Degen,giovannetti2011advances} harness nonclassical resources to substantially improve the performance of  positioning and timing~\cite{giovannetti2001quantum}, target detection~\cite{Lloyd2008,tan2008quantum,barzanjeh2015,zhang2015,zhuang2017,zhuang2017lidar}, digital reading~\cite{Qreading}, photometry~\cite{Photometry,Photometry2}, distributed sensing~\cite{zhuang2018distributed,proctor2018multiparameter,ge2018distributed,guo2020distributed,xia2019entangled,zhuang2020distributed}, and most prominently the Laser Interferometer Gravitational-wave Observatory (LIGO)~\cite{ligo,LIGO_2,tse2019_ligo}. There have been attempts to develop quantum-metrology protocols~\cite{RMP2016,mukamel2020roadmap} for spectroscopy~\cite{demtroder2013laser,levenson2012introduction,spec_RMP1,spec_RMP2,spec_RMP3}, an indispensable tool for science and industry. In this regard, the entangled NOON state~\cite{NOON1,NOON2} and its generalizations have been considered in interferometric estimation of loss and phase~\cite{dinani2016quantum}. However, NOON states are hard to generate and lack robustness against imperfections. As a more experimentally accessible approach, entangled photons produced by spontaneous parametric down conversion have been utilized for loss estimation~\cite{whittaker2017absorption,li2018enhanced,nair2018,kalashnikov2016infrared,paterova2018measurement,paterova2020hyperspectral}. In particular, Refs.~\cite{kalashnikov2016infrared,paterova2018measurement,paterova2020hyperspectral} reported nonlinear interferometric probing of mid-infrared absorption lines using visible photons. Despite the technical benefits, the quantum advantage over the classical schemes remains unclear.

\begin{figure}
    \centering
    \includegraphics[width=0.4\textwidth]{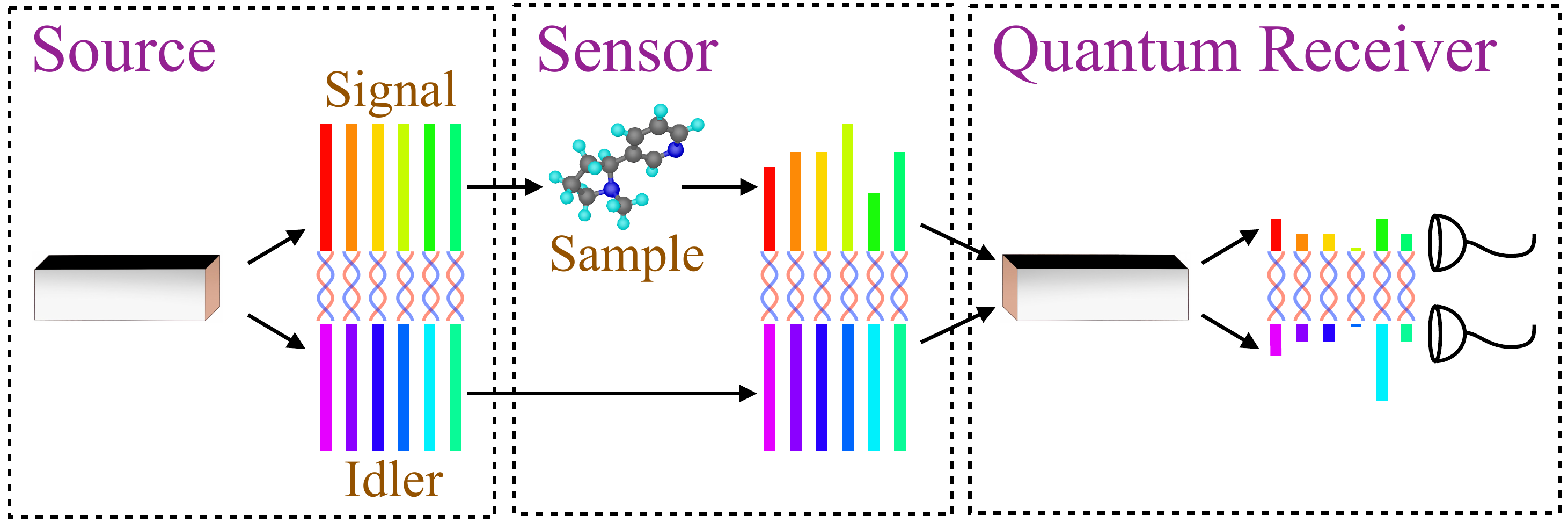}
    \caption{Diagram of the entanglement-assisted absorption spectroscopy. The source generates multi-chromatic entangled signal-idler pairs via a nonlinear process. The signals interact with the sample and then go through another nonlinear process jointly with the idlers at the quantum receiver. Photo detection extracts the absorption coefficients at different frequencies.
    \label{fig:schematic_concept}
    }
\end{figure}

In this work, we investigate entanglement-assisted absorption spectroscopy (EAAS), as an effective means to achieve a provable quantum advantage over all schemes using classical sources. As depicted in Fig.~\ref{fig:schematic_concept}, EAAS uses a source of multichromatic entangled signal-idler mode pairs from a nonlinear media, each being in a two-mode squeezed vacuum (TMSV) state and anti-correlated in the frequency domain. 
The signals with different frequencies interact with the sample and experience absorption differently, while the idlers are stored locally. Then an optical parametric amplifier (OPA) is applied on the return signal-idler pairs, followed by photodetection to classify samples among a plural of possibilities.

EAAS achieves a strict quantum advantage in the discrimination of arbitrary absorption patterns. Before addressing the general case, we begin with two basic models: absorption detection---the binary testing of a single absorption line at a specific frequency, and peak positioning---pinpoint a given number of absorption lines within a frequency spectrum. 
Then we consider the classification of several large organic molecules, and use real spectrum data~\cite{NIST} to simulate the performance against the optimum classical performance in `wine-tasting' and `drug-testing'. Let us remark that all components in our EAAS are off-the-shelf, and the quantum advantage is robust against excess noise and idler storage loss, making experimental implementations possible in the near-term.

{\em Pattern recognition on absorption spectra.--~}
\QZ{In absorption spectroscopy~\cite{demtroder2013laser}, each specific composition is associated with a unique absorption spectrum determined by measuring the transmissivities across the spectrum the input light. Therefore, the overall problem of composition identification can be formulated as a hypothesis testing between several known patterns of the frequency-dependent transmissivities, as formulated below.}

\QZ{
The multichromatic input light is decomposed into $m$ discrete frequency modes, denoted by the annihilation operators $\{a_\ell\}_{\ell=1}^m$. The input-output relation for each mode $a_\ell$ is modelled as a thermal loss channel $\calL^{\kappa_\ell,N_B}$~\cite{Weedbrook_2012} described by the Bogoliubov transformation 
\be 
a_\ell\to \sqrt{\kappa_\ell} a_\ell +\sqrt{1-\kappa_\ell} e_\ell,
\label{input_output}
\ee 
where $\kappa_\ell$ is the transmissivity and $e_\ell$ is a thermal mode with mean photon number $N_B/(1-\kappa_\ell)$ to model the environmental thermal noise, which is negligible ($N_B\sim 0$) at the optical wavelengths. However, to demonstrate the robustness of the quantum advantage, $N_B > 0$ is considered for generality. 
}

The pattern of the transmissivity coefficient $\{\kappa_\ell\}_{\ell=1}^m$ reveals the sample's absorption spectrum. We usually have prior information about the possible patterns, therefore the task is to discriminate between $H$ patterns, each described by a vector $\bm \kappa^{(h)}=\{\kappa_\ell^{(h)}\}_{\ell=1}^m$ of transmissivities, where $1\le h \le H$ is the index of the hypotheses \QZ{and $\ell$ is the frequency-mode index (see Ref.~\cite{supp} for a channel formulation)}. \QZ{In general, we allow $M$ repetitions of the probing attempt to make a decision.}

Before addressing the general pattern-recognition problem described above(see Fig.~\ref{fig:schematic_concept}), we consider two simplified problems: 
absorption detection
and peak positioning.


\QZ{In absorption detection, the goal is to determine whether absorption occurs at a single frequency mode ($m=1$), therefore there are $H=2$ hypotheses, with transmissivities $\kappa_B$ and $\kappa_T$ corresponding to the absence and the occurrence of absorption.} 
In peak positioning, one aims to pinpoint a single absorption peak (target) within $m$ frequencies, therefore we have $H=m$ possible patterns. \QZ{Each pattern $h$ has a single absorption peak with transmissivity $\kappa_T$ for frequency mode $a_h$ while all other frequency modes see a background transmissivity $\kappa_B$, i.e., $\kappa_{\ell}^{(h)}=\kappa_T$ if $\ell=h$, and $\kappa_B$ otherwise.}

The problem of absorption detection can be generalized to finding the positions of $k$ absorption peaks in a spectrum of $m$ frequencies, which we call `$k$-peak positioning'. In this more general problem, $k$ \QZ{targets with transmissivity $\kappa_T$} are hidden among $m-k$ backgrounds \QZ{of transmissivity $\kappa_B$}, so that we have a total of $H=C_m^k$ hypotheses, where $C_m^k$ is the binomial coefficient of $m$-choose-$k$. \QZ{Note that, while we consider these simple examples to introduce our results, our methodology applies to the recognition of arbitrary patterns, such as the complex molecules considered at the end of this paper.}

\begin{figure}
    \centering
    \includegraphics[width=0.4\textwidth]{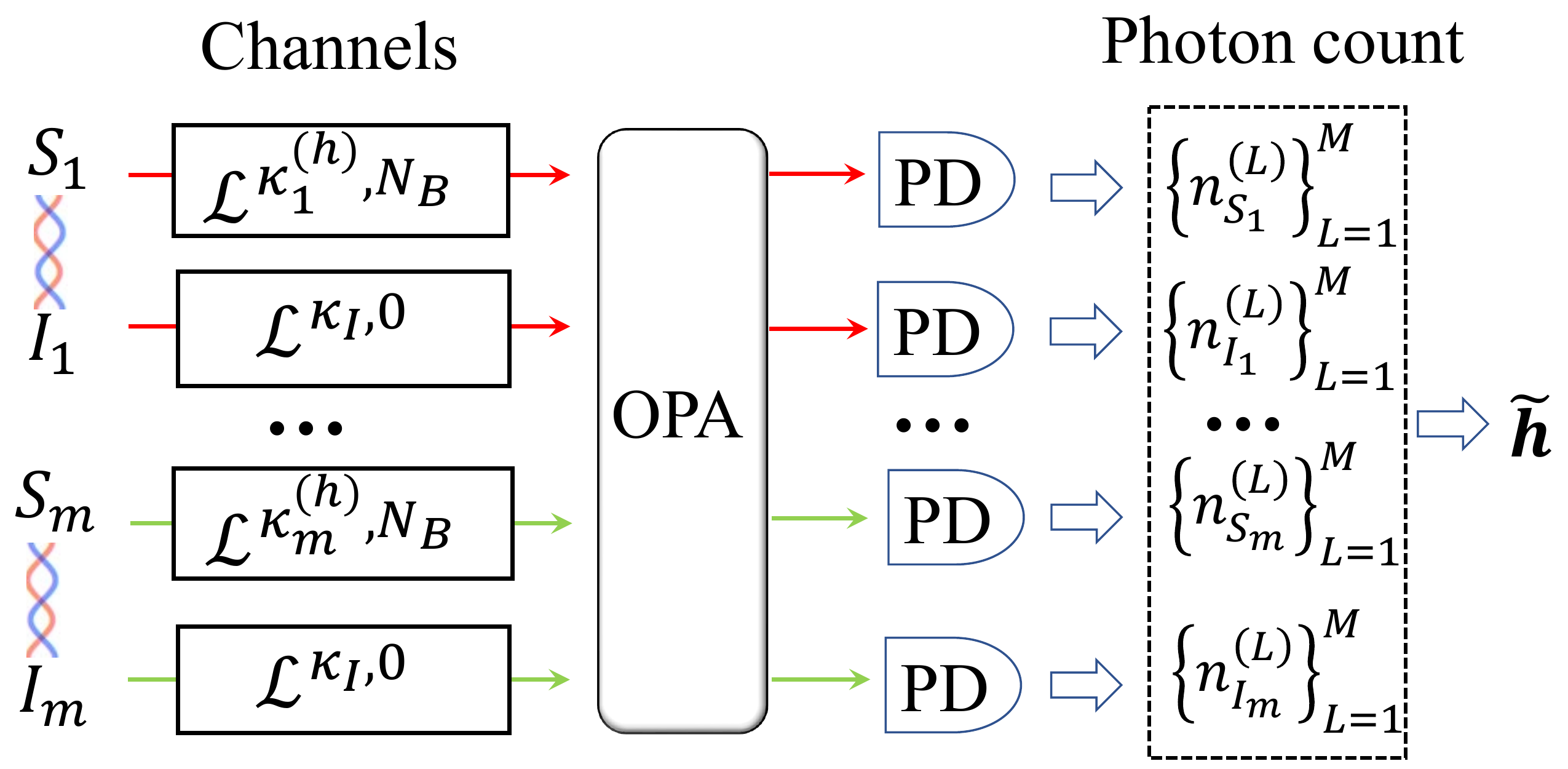}
    \caption{Schematic of receiver. Signal beams are irradiated over the sample, modeled by frequency-dependent transmissivities $\bm \kappa^{(h)}=\{\kappa_\ell^{(h)}\}_{\ell=1}^m$. The modulated beams go through a single optical parametric amplifier (OPA). Finally spectrally-resolving photodection (PD) offers a $2mM$-dimensional count, based on which the maximum-likelihood decision $\tilde{h}$ is made.
    \label{fig:PPMschematic}
    }
\end{figure}

{\em Classical lower bounds.---}~In a classical spectroscopy scheme, one sends an arbitrary mixture of coherent states as input state. Given $mMN_S$ mean total number of photons at the input, \QZ{where $N_S$ is the average mean photon number per frequency mode,} the minimum error probability affecting the discrimination between the ensemble of \QZ{patterns $\{\bm \kappa^{(h)}\}_{h=1}^H$} is lower bounded by
\begin{align}
&P_{C,m,LB}=\frac{2}{(H-1) H^3}\times
\label{LB_general}
\\
&\left[\min_{\{X_\ell\}} \sum_{h^\prime>h}   \exp\left({-\frac{1}{2}\sum_{\ell=1}^m (\sqrt{\kappa_\ell^{(h)}}-\sqrt{\kappa_\ell^{(h^\prime)}})^2 \nu_B X_\ell}\right)\right]^2, \nonumber 
\end{align}
where $\nu_B=1/(1+2N_B)$ and the minimization is under the energy constraint $\sum_{\ell=1}^m X_\ell \le mMN_S$ (See \cite{supp} for a proof). Applying Eq.~(\ref{LB_general}) to the absorption detection case, we obtain the lower bound $P_{C,1,LB}=e^{-\nu_BM N_S\left(\sqrt{\kappa_B}-\sqrt{\kappa_T}\right)^2 }/4$. In this case, a slightly improved bound can be obtained~\cite{Qreading}
\be
P_{C,1,LB}=\frac{1}{2}\left(1-\sqrt{1-e^{-\nu_BM N_S\left(\sqrt{\kappa_B}-\sqrt{\kappa_T}\right)^2 }}\right).
\label{C_Helstrom}
\ee
Specifying Eq.~(\ref{LB_general}) to the problem of $k$-peak positioning, one obtains~\cite{supp}
\be
P_{C,m,LB} = \frac{C_m^k-1}{2C_m^k} e^{-2w_{m,k}\nu_BM N_S(\sqrt{\kappa_B}-\sqrt{\kappa_T})^2},
\label{LB_Gaussi}
\ee
where $w_{m,k}={kC_{m-1}^k}/{(C_m^k-1)}$. The latter term is equal to $1$ for a single peak, and $w_{m,k}\simeq k(1-k/m)$ for $k$ peaks.
When $N_B=0$, the lower bound is tight in the error exponent for absorption detection and 1-peak positioning.

\begin{figure}
    \centering
    \includegraphics[width=0.4\textwidth]{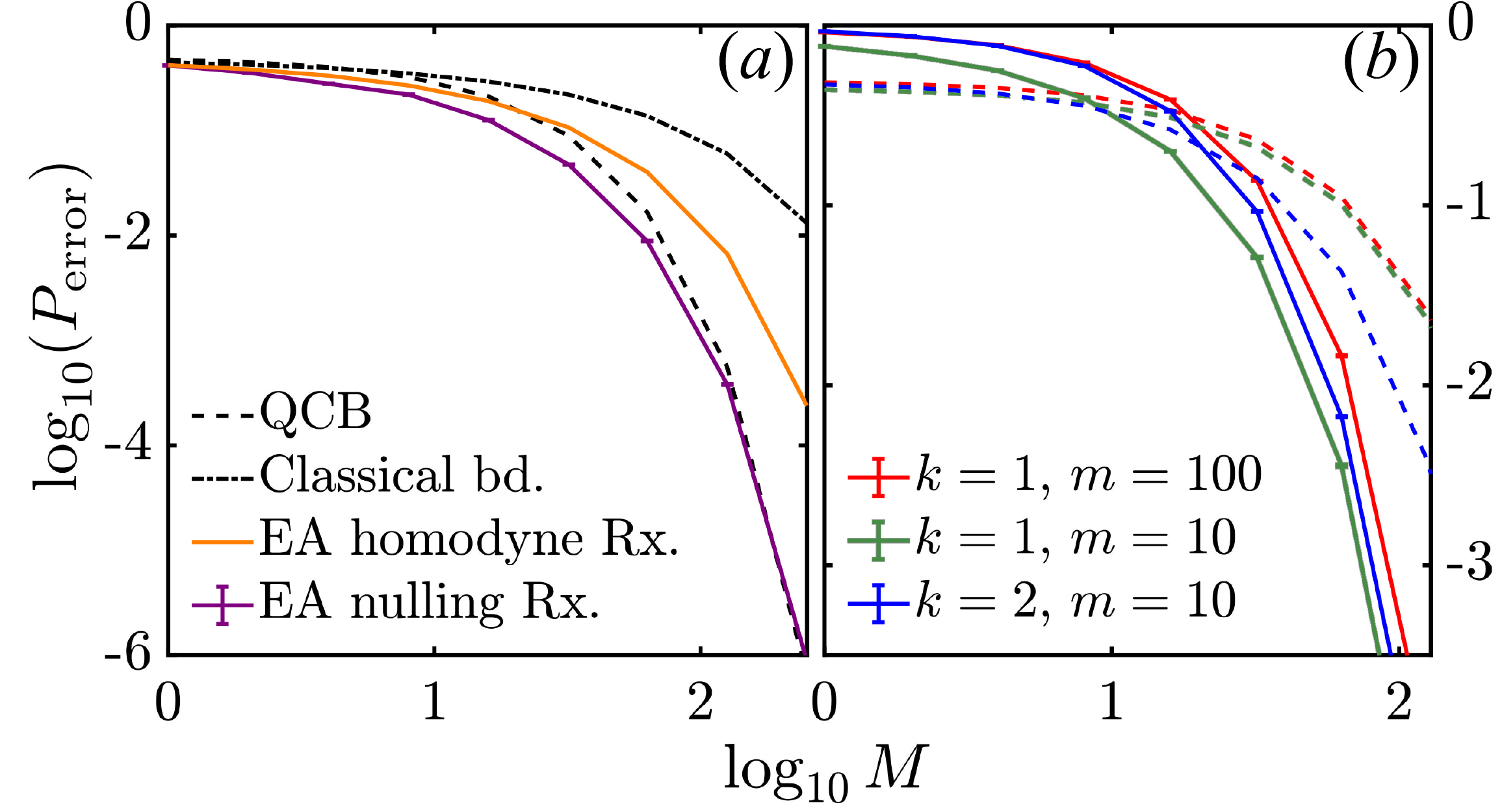}
    \label{fig:practical}
     \caption{Error rate versus number of probing modes with practical parameters $N_S=1$, $\kappa_T=0.75$, and $\kappa_B=0.95$. (a) Absorption detection. EA nulling receiver (solid orange) is compared with classical lower bound of Eq.~(\ref{C_Helstrom}) (dot-dashed black), QCB (dashed black), and the EA homodyne receiver (solid purple). (b) Peak positioning. Single-peak positioning (solid green) and double-peak positioning (solid blue) with $m=10$ frequency slots. Single-peak positioning with $m=100$ (solid red) provided as a reference. Classical lower bounds given by Eq.~(\ref{LB_Gaussi}) (dashed, accordingly colored).
     \label{fig:M}
     }
\end{figure}

{\em Entanglement-assisted strategy.--~}To achieve a quantum advantage, we exploit entanglement at the input, as given by $M$ copies of a TMSV state $\phi_{ME}$ for each signal-idler pair~\cite{supp}.
\QZ{Each idler mode is stored locally, with imperfections modeleld as a pure-loss channel $\calL^{\kappa_I,0}$ of transmissivity $\kappa_I$ (with a mode transformation $a_I\to \sqrt{\kappa_I} a_I+\sqrt{1-\kappa_I}v$ and $v$ being a vacuum mode); while the signal modes are sent to probe the patterns.}
For the special binary case of absorption detection, the error probability is bounded by the asymptotically-tight quantum Chernoff bound (QCB), which can be efficiently calculated~\cite{Pirandola2008,pirandola2011quantum} from the return Gaussian states
$
\Xi^{(T/B)}
$
\QZ{composed of $M$ identical copies of $\calL^{\kappa_{T/B},N_B}\otimes \calL^{\kappa_I,0} \left(\phi_{ME}\right)$.}
For the general pattern case, a simple tool like the QCB is missing and, for this reason, we need to design an explicit receiver that is able to show a quantum advantage.

{\em Entanglement-assisted receiver design.--~}
We begin our description of the receiver design with a simple case so as to provide its basic modus operandi. Consider the ideal case of $\kappa_B=\kappa_I=1$ and $N_B=0$. Then the returned state $\Xi^{(B)}=\phi_{ME}^{\otimes M}$ consists of $M$ copies of the ideal TMSV state (while $\Xi^{(T)}$ is mixed because $\kappa_T<1$). Suppose that we perform a two-mode squeezing (TMS) operation $\calS$ (via an OPA), that precisely anti-squeezes each TMSV state $\phi_{ME}$. 
Then we can `null' $\calS(\Xi^{(B)})$ to tensor products of vacuum, while $\calS(\Xi^{(T)})$ is non-vacuum. Therefore, a simple photon counting on all the signals and idlers after the TMS operation can identify the input state $\Xi^{(B)}$ if there is any photon count. Errors only occur if we obtain a zero count on $\calS(\Xi^{(T)})$: when this happens, we can only guess randomly, with an error probability $R_m$. \QZ{Note that this nulling strategy has been used in classical schemes~\cite{supp}, whose performance is bounded by Eqs.~\eqref{C_Helstrom} and~\eqref{LB_Gaussi}.} \QZ{OPA has also been utilized in quantum illumination~\cite{Guha2009}, however without exploiting correlations in the patterns.}
\begin{figure}
    \centering
    \includegraphics[width=0.4\textwidth]{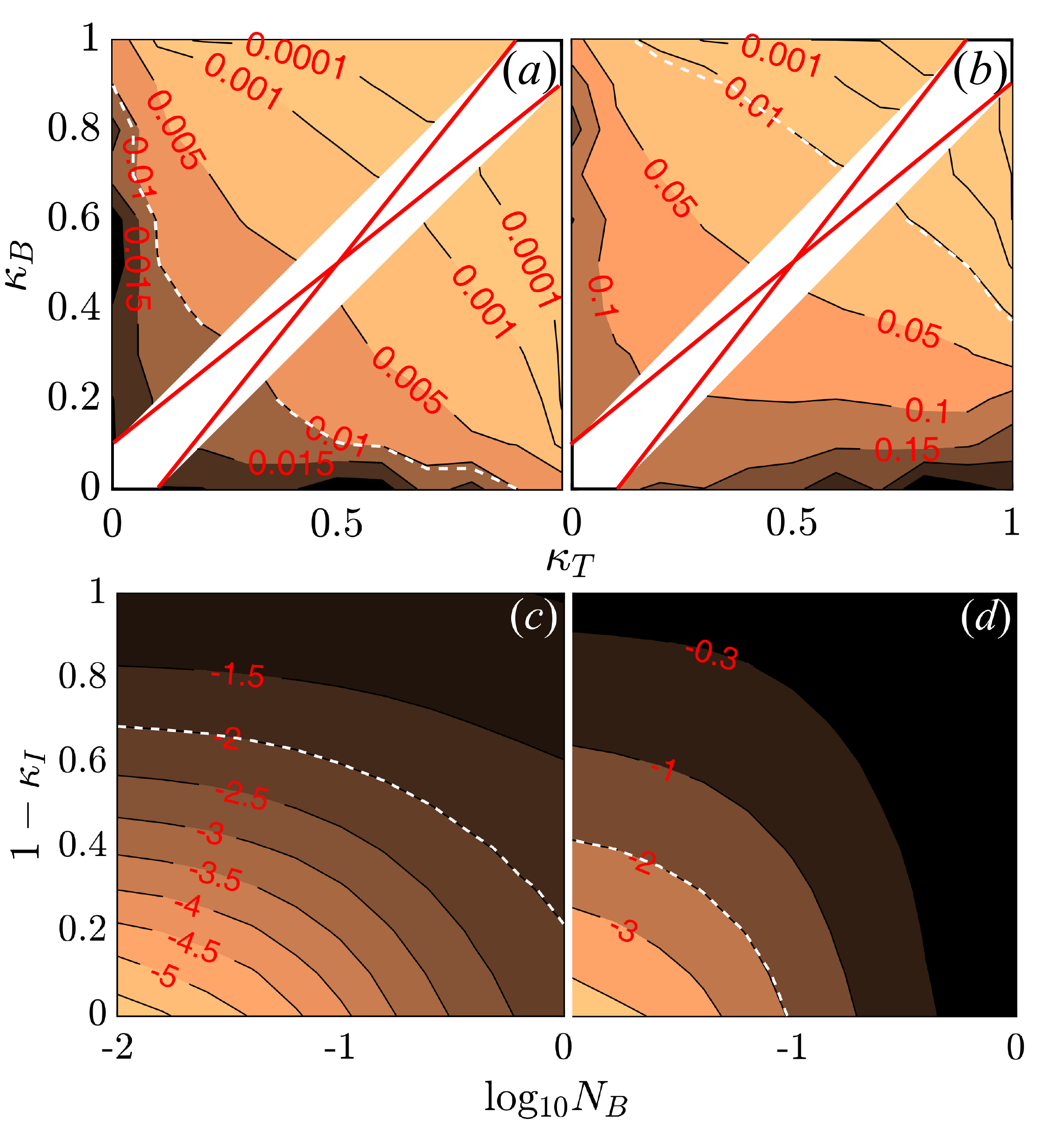}
     \caption{
     (a)(b)~Error probability of EAAS $P_{E,m}$ versus transmissivities $\kappa_{B}$ and $\kappa_{T}$; (c)(d) ${\rm\log}_{10}P_{E,m}$ versus idler loss $1-\kappa_I$ and thermal noise $N_B$ at fixed $\kappa_T=0.75$ and $\kappa_B=0.95$. absorption detection ($m=1$) in (a)(c) compared with single-peak positioning ($m=100$) in (b)(d). $M$ is chosen to fixed the classical lower bounds to $~0.01$~\cite{supp}. $N_S=1$ is assumed. Red-crossed diagonal region in (a)(b) represents the degenerate case $\kappa_B=\kappa_T$.
     \label{fig:contours_ideal_main}
     }
\end{figure}

Let us use a compact notation, where $m=1$ corresponds to absorption detection, for which $R_{1}=1/2$; and $m\ge2$ corresponds to single-peak positioning, with one copy of $\calS(\Xi^{(T)})$ among $m-1$ copies of $\calS(\Xi^{(B)})$, so that $R_{m}=\left(m-1\right)/m$. Accounting for the zero counts, the error probability for absorption detection ($m=1$) and single-peak positioning ($m \ge 2$) is given by~\cite{supp}
\be
P_{E,m}=R_m\left[\frac{1}{1+N_S(1-\sqrt{\kappa_T})}\right]^{2M}.
\label{eq:PPManalyt}
\ee
When $N_S\ll1$ and $M\gg1$, we have $P_{E,m}\simeq R_m\exp\left(-2MN_S\left(1-\sqrt{\kappa_T}\right)\right)$. Comparing this with the classical lower bounds in Eqs.~(\ref{C_Helstrom}) and~(\ref{LB_Gaussi}), we see that EAAS has an exponential advantage:
$ 
P_{E,m}/P_{C,m,LB}\simeq 2\exp\left(-MN_S\left(1-\kappa_T\right)\right)
$
for absorption detection, and
$
\simeq 2\exp\left(-2MN_S\left(\sqrt{\kappa_T}-\kappa_T\right)\right)
$
for single-peak positioning. In fact, Eq.~(\ref{eq:PPManalyt}) achieves the QCB~\cite{supp} and therefore it is optimal for absorption detection. 

\QZ{
The above receiver design, and the resulting entanglement advantage, can be generalized to cope with more complex spectrum patterns and the presence of noise and idler loss ($N_B>0,\kappa_I<1$), as described by the following strategy (see Fig.~\ref{fig:PPMschematic}): (i)~Apply TMS operation with gain $G_\ell$ to each of the return signal-idler pairs $a_{S\ell}^{\prime\prime},a_{I\ell}^{\prime\prime}$ to obtain new modes $a_{S\ell}=\sqrt{G_\ell}a_{S\ell}^{\prime\prime}-\sqrt{G_\ell-1}a_{S\ell}^\dagger$ and $a_{I\ell}=\sqrt{G}a_{I\ell}^{\prime\prime}-\sqrt{G-1}a_{S\ell}^\dagger$. (ii)~Perform photon counting measurement on all signal and idler modes $\{a_{S\ell}, a_{I\ell}\}$'s to obtain the results as two vectors $\bm n_S$ and $\bm n_I$. (iii)~Finally, apply maximum-likelihood (ML) decision rule, i.e., make the decision $\tilde{h}$ through
\be 
\tilde{h}=\arg\max_h P_m(\bm n_S,\bm n_I|h),
\label{h_MLE}
\ee 
where $P_m(\bm n_S,\bm n_I|h)$ is the conditional probability of obtaining the outcomes $\bm n_S,\bm n_I$ if the true hypothesis is $h$. 
}

To complete the description of our receiver, we need to determine the gain $G_\ell$'s and specify the conditional probabilities. Let us begin with the cases of absorption detection and peak positioning, where we adopt uniform gain $G_\ell=G$; the ideal situation is to get a quantum state close to vacuum; however, if $\kappa_B<1$, it is only possible to reduce the signal part of $\Xi^{(B)}$ to vacuum, by choosing $
G=1+{N_S\kappa_B}/{(1+N_S(1-\kappa_B))}
$.
In the presence of noise $N_B>0$ and idler loss $\sqrt{1-\kappa_I}>0$, `nulling' to vacuum is not possible but the same choice of gain still provides an appreciable advantage over classical schemes. 
\QZ{
For general patterns, due to the absence of symmetry we consider optimization over the gain $G_\ell$'s at different frequency modes. Moreover, as some frequency windows may contain more essential information about the hypotheses, we also allow the optimization over the energy distribution $\{N_{S\ell}\}$ of the TMSV in different frequency modes. In these cases, although the `nulling' decision rule does not apply, the ML decision rule in Eq.~\eqref{h_MLE} still leads to an advantage~\cite{supp}.
}

Now let us compute the conditional probabilities. 
With $M$ identical repetitions, the probability of obtaining the $mM$-dimensional measurement results $\bm n_S=\{n_{S_L,k}\}_{L=1,k=1}^{M,m}$ and $\bm n_I=\{n_{I_L,k}\}_{L=1,k=1}^{M,m}$, conditioned on pattern $h$, is 
\begin{align}
&P_m(\bm n_S,\bm n_I|h)
=\prod_{L=1}^{M}\prod_{\ell=1}^m P(n_{S_{L,\ell}},n_{I_{L,\ell}}|\kappa_\ell^{(h)},G_\ell,N_{S\ell}),
\label{condtional_prob}
\end{align}
where each term is a function of the subsystem transmissivity $\kappa_\ell^{(h)}$, the TMSV source energy $N_{S\ell}$ and the gain choice $G_\ell$~\cite{supp}.

With all these theoretical elements in our hands, we can numerically evaluate the error probability $P_{E,m}$ for the problems of absorption detection, peak positioning and general spectrum recognition via Monte Carlo simulations~\cite{supp}. Although we consider equal priors for simplicity, our ML decision can generally be applied to arbitrary prior probabilities for the patterns.

\begin{figure}
    \centering
    \includegraphics[width=0.5\textwidth]{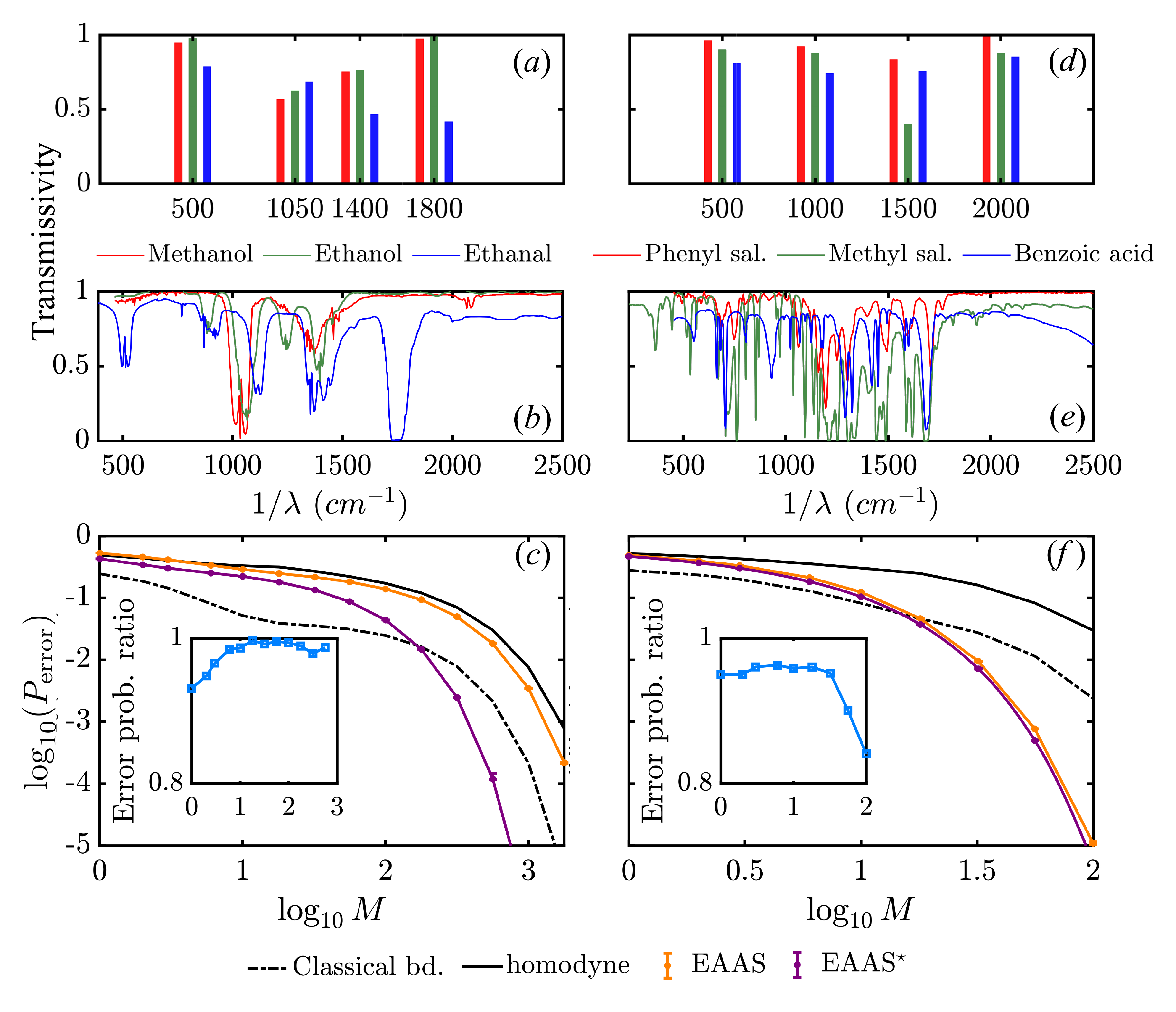}
    \caption{Identification of $H=3$ molecules with $m=4$ frequency slots, for wine-tasting in (a)(b)(c) and drug-testing in (d)(e)(f). (a)(d) are $m=4$ sampled discrete spectra on the FTIR spectra in (b)(e). (c) and (f) show the logarithmic error rate of EAAS, with $N_S = 1$. \QZ{`EAAS' assumes a uniform distribution of photons at the input modes (solid orange), while numerically optimized 
   energy distribution presented in `EAAS$^\star$' (solid purple).} The classical lower bound (dot-dashed black) and homodyne detection (solid black). \QZ{Insets of (c) and (f): error probability ratio of EAAS with gain optimization after energy-distribution optimization and EAAS with merely energy-distribution optimization.}
    \label{fig:molecule}
    }
\end{figure}

{\em Detecting and positioning absorption peaks.--~}In order to investigate the problems of absorption detection and peak positioning, we assume a background transmissivity $\kappa_B=0.95$ and a target transmissivity $\kappa_T=0.75$. In particular, we study their error probabilities in terms of the number of modes $M$. For absorption detection, 
Fig.~\ref{fig:M}(a) shows that our EA nulling receiver asymptotically achieves the QCB~\cite{Qreading}, outperforming both the best known receiver, the EA homodyne receiver~\cite{supp,pirandola2011quantum}, and the classical lower bound of Eq.~(\ref{C_Helstrom}).
In fact, we can verify that our receiver can asymptotically saturate the QCB for absorption-detection with general choices of $\kappa_B$ and $\kappa_T$~\cite{supp}. For peak positioning, as shown in Fig.~\ref{fig:M}(b), our EA receiver is able to outperform the classical lower bound of Eq.~(\ref{LB_Gaussi}) by orders of magnitude.

In a practical scenario, we are interested in how much EAAS can enhance the performance, when classical schemes fail to perform well. To showcase the advantage, in Fig.~\ref{fig:contours_ideal_main}, we fix the classical lower bounds in Eqs.~(\ref{C_Helstrom}) and~(\ref{LB_Gaussi}) to be $0.01$ and plot the error probability $P_{E,m}$ of EAAS. We start with tuning the transmissivities $\kappa_B$ and $\kappa_T$ in Fig.~\ref{fig:contours_ideal_main}(a)(b). Then we fix $\kappa_T=0.75$ and $\kappa_B=0.95$ and study how the quantum advantage varies with idler loss $1-\kappa_I$ and noise $N_B$ in Fig.~\ref{fig:contours_ideal_main}(c)(d). The white dashed lines divide the parameter space with/without quantum advantage. We can see that the advantage is remarkable, and also survives for a large range of parameters, especially when $\kappa_B\simeq 1$ as in practice. The robustness of the advantages to imperfections reveals a clear possibility for a near-term experimental demonstration. See \cite{supp} for more parameter settings. 


{\em General spectrum recognition.--~} EAAS can also identify actual molecules, each of which is associated with a unique absorption spectrum. 
As a taste of flavor, we begin with `wine-tasting'---where one discriminates three common alcohol-like liquids. Methanol could be lethal if mistaken for ethanol (alcohol). Meanwhile, the alcohol, as time goes by, will be dehydrogenated to ethanal, whose concentration provides the age of a vintage~\cite{yu2008prediction}. To consider larger molecules, the second example, `drug-testing', involves three drugs: phenyl salicylate, methyl salicylate, and benzoic acid. In both examples, a nondestructive testing method is preferred, as we shall conduct with the extremely weak quantum light source. The transmissities are taken from real Fourier-transform infrared (FTIR) spectra~\cite{NIST}. These spectra are sampled by averaging them within each of $m=4$ frequency slots~\cite{supp}.


As the classical benchmark, we calculate the ultimate lower bound using Eq.~(\ref{LB_general}) and the performance of a homodyne receiver on coherent-state input with the same energy distribution (distribution of mean photon number over frequency modes) optimized in Eq.~(\ref{LB_general}). Fig.~\ref{fig:molecule} shows that EAAS with uniform energy distribution and $G=1$ (orange) outperforms the homodyne receiver (black solid) in both cases. Then, in drug-testing, EAAS beats the classical lower bound by orders of magnitude, while, in wine-tasting, this advantage is less pronounced. \QZ{This is mainly due to the classical lower bound being not tight, and uniform energy being sub-optimum, as we see EAAS with energy optimization (purple) enables much better advantages. Although gain optimization only leads to a slight advantage over the energy-optimized EAAS, as evident in the inset plots of Fig.~\ref{fig:molecule}(c)(f). In the noisy case, it enables a much better enhancement~\cite{supp}.
}

\QZ{
Now we address phase noise common in experiments.
Phase tracking can typically eliminate the time-invariant phase noise, so the above results directly hold; when phase tracking is not possible, we can model the phase noise by adding $a_\ell\to e^{i\theta_\ell}a_\ell$ in Eq.~\eqref{input_output}. The random phase $\theta_\ell$ clearly complicates the problem. However, if we choose uniform $G=1$ (i.e., not applying OPA before photodetection), the same results of the orange curves in Fig.~\ref{fig:molecule}(c)(f)  hold, and the classical performance can only be worse than the current benchmarks (black). Thus, the quantum advantage sustains. 
}

{\em Conclusion.---}
We have devised a near-term feasible EAAS scheme that outperforms any classical strategy in determining the presence and position of spectral absorption peaks. The EAAS scheme saturates the QCB in binary detection of a single absorption line and offers orders-of-magnitude advantage in error probability in the discrimination of sampled spectra of molecules even in the presence of experimental nonidealities.



\bigskip
{\em Acknowledgements.---}
This research is supported by Defense Advanced Research Projects Agency (DARPA) under Young Faculty Award (YFA) Grant No. N660012014029 and University of Arizona. S.P. acknowledges funding from EU Horizon 2020 Research and Innovation Action under grant agreement No. 862644 (Quantum Readout Techniques and Technologies, QUARTET). Z.Z. is supported by the Office of Naval Research Grant No.~N00014-19-1-2190.

\newpage

\pagebreak
\begin{center}
\begin{widetext}
\textbf{\large Supplemental Materials: Entanglement-Assisted Absorption Spectroscopy}
\end{widetext}
\end{center}

\makeatletter
\renewcommand{\theequation}{S\arabic{equation}}
\renewcommand{\thefigure}{S\arabic{figure}}
\renewcommand{\bibnumfmt}[1]{[S#1]}
\renewcommand{\citenumfont}[1]{S#1}

\section{Channel set-up}
In the main paper, we mainly utilize input-output relations to describe the quantum channels, here we make the quantum channel notations explicit.

The overall $H$ patterns being discriminated can each be modelled as quantum channels $\{\calJ^{(\bm k^{(h)})}\}_{h=1}^H$ act on $m$ subsystems. They are given by
$ 
\calJ^{(\bm k^{(h)})}=\otimes_{\ell=1}^m \Phi^{(h)}_{S_\ell},
$ 
where each sub-channel $\Phi^{(h)}_{S_\ell}\equiv \left(\calL^{\kappa_\ell^{(h)},N_B}\right)^{\otimes M}$ is associated with $M$ probings of subsystem $S_\ell$ and $\calL^{\kappa,N}$ is a thermal loss channel with transmissivity $\kappa$, noise $N$.

In the absorption detection case, $H=2$, $m=1$, and we denote the two channels as $\Phi^{(T/B)}=\left(\calL^{\kappa_{B/T}, N_B}\right)^{\otimes M}$. The transmissivities $\kappa_B$ and $\kappa_T$ correspond to the absence (background channel $\Phi^{(B)}$) and the occurrence (target channel $\Phi^{(T)}$) of absorption.

In the peak positioning case, we have $H=m$ possible global channels, where $\calJ^{(\bm k^{(h)})}$ has a target channel $\Phi^{(T)}$ at subsystem $S_h$ while the rest are transparent backgrounds $\Phi^{(B)}$. The corresponding pattern is therefore described by $\kappa_{\ell}^{(h)}=\kappa_T$ if $\ell=h$, and $\kappa_B$ otherwise. 

In the general pattern case, we consider discrimination between the ensemble of channels $\{\calJ^{(\bm k^{(h)})}\}_{h=1}^H$.

When we introduce entanglement assistance, each sub-channel is extended to $\Phi^{(h)}_{S_\ell}\otimes \calD^{\kappa_I}_{I_\ell}$, where $\calD^{\kappa_I}=\left(\calL^{\kappa_I,0}\right)^{\otimes M}$ and $\calL^{\kappa_I,0}$ is a noiseless lossy channel modelling the imperfections on each idler system $I_\ell$.
Then the return state for each sub-channel is
$ 
\Xi^{(h)}_{S_\ell}= \Phi^{(h)}_{S_\ell}\otimes \calD^{\kappa_I}_{I_\ell}\left(\phi_{ME}^{\otimes M}\right)
$. For the case of target/background channels, the return states
\be 
\Xi^{(T/B)}= \Phi^{(T/B)}\otimes \calD^{\kappa_I}\left(\phi_{ME}^{\otimes M}\right).
\ee 

\section{Covariance matrix and two-mode photon statistics}
\label{sec:photonstat}

\subsection{Covariance matrix derivation}

First, we briefly introduce the notion of Gaussian states~\cite{Weedbrook_2012}, whose Wigner functions have a Gaussian shape. An $n$-mode Gaussian state $ {\rho}$ comprising modes $a_k, 1\le k \le n$, is fully characterized by the mean and the covariances of real quadrature field operators $q_k= {a}_k+ {a}_k^\dagger, p_k=i\left( {a}_k^\dagger- {a}_k\right)$. Formally, we can define a real $2n$-dim vector of operators
${ {\bm x}}=\left(q_1,p_1,\cdots, q_n,p_n\right)$, then the mean $\bar{\bm x}=\expval{ {\bm x}}_{ {\rho}}$ and the elements of the $2n$-by-$2n$ covariance matrix are given by
\be
{\bm \Lambda}_{ij}=\frac{1}{2} \expval{\{ {x}_i-\bar{x}_i, {x}_j-\bar{x}_j\}}_{ {\rho}},
\ee
where $\{,\}$ is the anticommutator and $\expval{ {A}}_{ {\rho}}={\rm Tr} \left( {A} {\rho}\right)$.

An important example of Gaussian state is TMSV, given by the wave-function
\be
\phi_{ME}=\sum_{n=0}^\infty \sqrt{\frac{N_S^n}{(N_S+1)^{n+1}}}\ket{n}_{S^\prime}\ket{n}_{I^\prime},
\label{wf_TMSV}
\ee 
where $\ket{n}$ is the number state.
From the above wave-function, we can obtain the covariance matrix of a TMSV as
\begin{align}
& 
{\mathbf{{\mathbf{\Lambda}}}}_{\rm TMSV} =
\left(
\begin{array}{cccc}
(2N_S+1) {\mathbf I}&2C_0{\mathbf Z}\\
2C_0 {\mathbf Z}&(2N_S+1){\mathbf I}
\end{array} 
\right),
&
\label{cov_TMSV}
\end{align}
where ${\mathbf I}$, ${\mathbf Z}$ are two-by-two Pauli matrices, and $C_0=\sqrt{N_S\left(N_S+1\right)}$ is the amplitude of the phase-sensitive cross correlation. 

We utilize multiple copies of the signal-diler pair $\{a_S^\prime,a_I^\prime\}$ in a TMSV state to probe the sample. To begin with, we consider the case with no phase noise, where each signal goes through the channel $\calL^{\kappa_S, N_B}$, giving the output
\be 
a_S^{\prime\prime}=\sqrt{\kappa_S}a_S^\prime+\sqrt{1-\kappa_S}e,
\ee 
where $e$ mode is in a thermal state with mean photon number $N_B/(1-\kappa_S)$. While the idler mode $a_I^\prime$ goes through a pure loss channel $\calL^{\kappa_I,0}$.
\be 
a_I^{\prime\prime}=\sqrt{\kappa_I} a_I^\prime+\sqrt{1-\kappa_I}v,
\ee 
where the environment mode $v$ is in vacuum state.
The state $\rho_{{\mathbf{{\mathbf{\Lambda}}}}_0\left({\kappa_S}\right)}$ of each of the signal-idler pair $\{a_S^{\prime\prime},a_I^{\prime\prime}\}$ has the covariance matrix
\begin{align}
& 
{\mathbf{{\mathbf{\Lambda}}}}_0\left({\kappa_S}\right) =
\left(
\begin{array}{cccc}
(2(\kappa_S N_S+N_B)+1) {\mathbf I}&2C_p{\mathbf Z}\\
2\kappa_S\kappa_IC_0 {\mathbf Z}&(2\kappa_I N_S+1){\mathbf I}
\end{array} 
\right),
&
\label{cov_TMSV_return}
\end{align}
with the signal-idler cross correlation $C_p=\sqrt{\kappa_S\kappa_I N_S(1 + N_S)}$.

In the case of absorption detection, we have $\kappa_S=\kappa_B$ or $\kappa_S=\kappa_T$.
With each Gaussian mode-pair of the return state specified by Eq.~(\ref{cov_TMSV_return}), we can obtain the QCB on the error probability of the binary hypothesis testing through methods in Ref.~\cite{Pirandola2008},
\be 
P_E\leq P_{QCB}=\tilde Q_s^M/2,
\ee
where $\tilde Q_s={\rm inf}_{0\leq s\leq 1}{\rm Tr}\rho_{{\mathbf{{\mathbf{\Lambda}}}}_0\left({\kappa_B}\right)}^s\rho_{{\mathbf{{\mathbf{\Lambda}}}}_0\left({\kappa_T}\right)}^{1-s}$. 
The QCB is asymptotically tight as the mode number $M\to \infty$ and can be efficiently calculated in our case~\cite{Pirandola2008,pirandola2011quantum}.

On the receiver side, we apply a two-mode squeezing process, parametrized by the gain $G\ge 1$, to obtain returned signal-idler mode pairs $\{a_S, a_I\}$ 
\begin{align}
a_S=\sqrt{G}a_S^{\prime\prime}-\sqrt{G-1}a_I^\dagger,
\\
a_I=\sqrt{G}a_I^{\prime\prime}-\sqrt{G-1}a_S^\dagger.
\end{align}
They are in a Gaussian state with covariance matrix
\be
{\mathbf{{\mathbf{\Lambda}}}}\left({\kappa_S}\right)=
\left(
\begin{array}{cccc}
E {\mathbf I}&C{\mathbf Z}\\
C {\mathbf Z}&S{\mathbf I}
\end{array} 
\right).
\label{cov_mat}
\ee
Denote $G=1+N_S^{\prime}$ with the effective photon number $N_S^{\prime}$, we have the variances of signal and idler 
\begin{align}
&E=1+2 \kappa_S N_S-4C_p\sqrt{ N_S^{\prime}(1 + N_S^{\prime})} 
\nonumber
\\
&+ 2N_S^{\prime}(1+\kappa_I N_S+ \kappa_S N_S)+ 
2N_B(1+N_S^{\prime}), 
\\
&S=1+2\kappa_I N_S-4C_p\sqrt{N_S^{\prime}(1 + N_S^{\prime})} 
\nonumber
\\
&+ 2N_S^{\prime}(1+\kappa_I N_S+ \kappa_S N_S)+
2N_BN_S^{\prime},
\end{align}
and the signal-idler correlation
\begin{align}
&C= 2\sqrt{N_S^{\prime}(1+N_S^{\prime})}(1+\kappa_I N_S+ \kappa_S N_S)
\nonumber
\\
&-2(1+2N_S^{\prime})C_p+2N_B\sqrt{(1+N_S^{\prime})N_S^{\prime}}.  
\end{align} 

When $\kappa_I=1,N_B=0$, suppose we choose
\be
N_{S0}^{\prime}=\frac{N_S\kappa_B}{1+N_S(1-\kappa_B)},
\label{eq:G0}
\ee
as given in the main paper, we have
\be
{\mathbf{{\mathbf{\Lambda}}}}\left({\kappa_B}\right)=
\left(
\begin{array}{cccc}
{\mathbf I}&{\mathbf 0}\\
{\mathbf 0}&\left(1+2\left(1-\kappa_B\right)N_S\right){\mathbf I}
\end{array} 
\right).
\label{cov_mat_kB}
\ee
This means that the signal mode becomes vacuum under this choice of the gain. We will choose this gain in the absorption detection and peak positioning cases, with the exception in Sec.~\ref{app:opt_gain}, in which we optimize the gain. For the general pattern recognition for molecules, we will consider both uniform gain $G=1$ and consider optimization over gain.

\subsection{Photon number statistics}
We consider photon counting on each signal-idler pair $\{a_S, a_I\}$ in a Gaussian state with the covariance matrix in the form of Eq.~(\ref{cov_mat}), which has zero phase-insensitive cross-correlation $\expval{a_S a_I^\dagger}=0$ and non-zero phase sensitive cross-correlation $\expval{a_S a_I}\neq 0$. The corresponding joint probability of obtaining results $n_S,n_I$ is given by
\bal
P(n_S,n_I)=& -4 F_R(1+n_S,1+n_I,1,\frac{4C^2}{XY})
\\
&\times \frac{(-1+C^2+E+S-E S)^{1+n_S +n_I}}{X^{1+n_S}Y^{1+n_I}},
\label{eq:prob}
\eal
where $F_R$ is the regularized hypergeometric function and 
\begin{align}
X&=1+C^2+E-(1+E)S,
\\
Y&=C^2-(E-1)(S+1).
\end{align}
Moreover, for states with covariance matrix equal to Eq.~(\ref{cov_mat}) up to a phase rotation on any mode, the photon number statistics is also given by Eq.~(\ref{eq:prob}).  

In our hypothesis testing protocols, $P(n_S,n_I)$ is fully determined by $N_S, N_S^{\prime},  \kappa_S, \sqrt{1-\kappa_I}, N_B$, where the last two $\sqrt{1-\kappa_I}, N_B$ are the fixed environment parameters, idler loss and thermal noise. Given signal mean photon number $N_S$, our receiver, parametrized by the gain $G=1+N_S^{\prime}$, is to jointly discriminate the photon count distributions per slot $\{P(\cdot,\cdot|\kappa_S,G,N_S)\}$ associated with different signal channel transmissivity patterns, conditioned on the same parameter setting $N_S^{\prime}, \kappa_I, N_B$. Here we make the dependence of Eq.~(\ref{eq:prob}) on $\kappa_S,G,N_S$ explicit.

\subsection{Conditional probabilities}
In the main paper, we give the conditional probability of obtaining the $mM$-dimensional measurement results $\bm n_S=\{n_{S_L,k}\}_{L=1,k=1}^{M,m}$ and $\bm n_I=\{n_{I_L,k}\}_{L=1,k=1}^{M,m}$ conditioned on pattern $h$ in Eq.~(\ref{condtional_prob}) of the main paper. Here we also print it for reference
\begin{align}
&P_{m}(\bm n_S,\bm n_I|h)
=\prod_{L=1}^{M}\prod_{\ell=1}^m P(n_{S_{L,\ell}},n_{I_{L,\ell}}|\kappa_\ell^{(h)},G_\ell,N_{S\ell}).
\label{condtional_prob_app}
\end{align}
Each term
$P(n_{S_{L,\ell}},n_{I_{L,\ell}}|\kappa_\ell^{(h)},G_\ell,N_{S\ell})$ is given in Eq.~(\ref{eq:prob}), with the transmissivity $\kappa_S=\kappa_\ell^{(h)}$, also the gain distribution $G_\ell=1+N_{S\ell}^\prime$ and the energy distribution $N_{S\ell}$ optimized probalistically. For the scenarios with strong symmetry, e.g. the peak positioning case, the optimal gain and energy distributions may be trivially uniform due to the symmetry.

For the absorption detection case, $m=1$ and $H=2$, the two hypotheses are $h=\{T,B\}$. Denote $M$-dimensional vectors of photon count $\bm n_S=\{n_{S_L}\}_{L=1}^M,\bm n_I=\{n_{I_L}\}_{L=1}^M$. Furthermore, with total energy being limited, the degree of freedom of $N_S$ is frozen. Eq.~(\ref{condtional_prob_app}) reduces to
\be 
P_m(\bm n_S,\bm n_I|h)=\prod_{L=1}^{M}P(n_{S_L},n_{I_L}|\kappa_h,G),
\ee
Similarly, for the single-peak positioning case, $H=m>1$, each hypothesis $h\in[1,m]$ corresponds to the position of the target channel. Noting the symmetry in this scenario, we choose the uniform energy and gain distributions. Still, the degree of freedom $N_S$ is frozen and that of gain is limited to one. Eq.~(\ref{condtional_prob_app}) reduces to
\begin{align} 
&P_{m}(\bm n_S,\bm n_I|h)
\nonumber
\\
&=\prod_{L=1}^{M}\![P(n_{S_{L,h}},n_{I_{L,h}}|\kappa_T,G)\cdot\prod_{l\not=h}P(n_{S_{L,l}},n_{I_{L,l}}|\kappa_B,G)].
\end{align}
In general, one can tune the energy and gain distributions among the different slots to approach the optimal performance.

\subsection{Limiting cases of the photon statistics}

To enable numerical simulation, we also need to deal with term-wise divergence in Eq.~(\ref{eq:prob}), e.g. the hypergeometric function $F_R$ could diverge. Since the probability is normalized, all divergences are in fact cancelled out by pairing infinite terms with infinitesimal terms. This must be done in an analytical way before numerical calculations.

The divergence comes from special values of $z=4C^2/XY$, as summarized in the following two cases:

\subsubsection{Case 1: $z \to \infty$}
This can be caused by $X\to 0$ or $Y\to0$.
This only happens when all noise are zero.
 Then
\bal
&\lim_{Y\to0} F_R(1+n_S,1+n_I,1,\frac{4C^2}{XY})/Y^{1+n_I}=\\
&\begin{cases}
      C^{-2(n_I+1)} (-4)^{-1 - n_I} X^{1+n_I}(-1)^{n_S}\binom{n_I}{n_S}, & \text{if}\ n_I\geq n_S \\
      0, & \text{otherwise}
    \end{cases}
\eal
Hence 
\bal
&P(n_S,n_I)=(-1+C^2+E+S-E S)^{1 + n_S + n_I}\\
&(-4)^{-n_I} (-1)^{n_S} C^{-2 (1 + n_I)} X^{n_I-n_S}
 \binom{n_I}{n_S}
 \label{eq:probY=0}
\eal
for $n_I\geq n_S$ or $0$ otherwise.

Two-mode squeezed vacuum is among this case, by taking $E,S=1+2N_S$, $C=2\sqrt{N_S(N_S+1)}$, we find $X=Y=0$, which makes the above expression non-zero only when $n_I=n_S$, and indeed we have $P(n_S,n_I)=N_S^{n_S}/(1+N_S)^{n_S+1} \delta_{n_S-n_I}$.

A special scenario of case 1 happens when $z=\frac{0}{0}$.
At this moment $X$ or $Y$ is zero, combined with $C=0$. To avoid singularity $C^{-2(1+n_I)}$ in numerical calculation, we take the $C\to 0$ limit with Eq.~(\ref{eq:probY=0}), which yields
\be
P(n_S,n_I)=
\begin{cases}
     2 (E-1)^{n_S} (1 + E)^{-1 - n_S}, & \text{if}\ n_I=0,\\
      0, & \text{otherwise},
    \end{cases}
\ee
for $X=0$  and 
\be
P(n_S,n_I)=
\begin{cases}
     2  (S-1)^{n_I} (1 + S)^{-1 - n_I}, & \text{if}\ n_S=0\\
      0, & \text{otherwise}.
    \end{cases}
\ee
for $Y=0$. Note that first case corresponds to signal mode thermal statistics 
\be 
P(n_S)=\frac{\overline N_S^{n_S}}{(1+\overline N_S)^{n_S+1}},
\ee 
with mean photon number $\overline N_S=(-1+E)/2$; the second case corresponds to idler mode thermal statistics 
\be 
P(n_I)=\frac{\overline N_I^{n_I}}{(1+\overline N_I)^{n_I+1}},
\ee 
with mean photon number $\overline N_I=(-1+S)/2$.

\subsubsection{Case 2: $z=4C^2/XY\to 1$}
In this case $F_R$ is infinite. Indeed it always comes with $-1+C^2+E+S-E S\to 0$ which cancels the singularity. We have
\bal
&P(n_S,n_I)=\left(\frac{E+S}{(E-1)(S-1)}\right)^{-1-n_S-n_I}(-1)^{-n_S-n_I}\times \\
&2^{3+n_S+n_I}\frac{(2-2E)^{-1-n_I}(2-2S)^{-1-n_S}(n_S+n_I)!}{n_S!n_I!}.
\eal

\section{Entangle-assisted homodyne receiver}
\label{app:pirandola's}
We give a brief summary of the best known receiver design for quantum reading, a Bell-measurement receiver proposed in Ref.~\cite{Qreading}, which we utilized to benchmark in Fig.~\ref{fig:M}(a) of the main paper. The returned signal-idler pair travels through a beamsplitter, which yields output modes $a_+=(a_S+a_I)/\sqrt{2}$, $a_-=(a_S-a_I)/\sqrt{2}$. Then homodyne measurements are operated on the two quadratures $p_+=(a_+-a_+^\dagger)/i, q_-=(a_-+a_-^\dagger)$, which share the same variance $V_h=\sigma^2_h=\expval{p_+^2}=\expval{q_-^2}=1 + N_S + N_S \kappa^{(h)} - 2 \sqrt{N_S (1 + N_S) \kappa^{(h)}}$, $h=\{T,B\}$. With a large identical mode number $M$, the receiver design constructs $\chi^2$ test variable $\theta=\sum_{L=1}^M p_{+,L}^2+q_{-,L}^2$, which comes with the probability density function 
\be 
P_{M}(\theta|h)=\frac{\theta^{M-1}\exp\left(-\frac{\theta}{2V_h}\right)}{(2V_h)^M\Gamma(M)},
\ee
where $\Gamma(\cdot)$ is the Gamma function. Under maximum-likelihood decision, the error rate is $(1-|F_T-F_B|)/2$, given the cumulative distribution function (cdf) of the chi-square distribution $F_h=\Gamma(M,\frac{t}{2V_h})/\Gamma(M)$ with the threshold 
$t= 2M \sigma^2_T \sigma^2_B\log(\sigma^2_T/\sigma^2_B)/(\sigma^2_T-\sigma^2_B)$. Here $\Gamma(\cdot,\cdot)$ is the incomplete Gamma function

Generally, the Bell receiver, extracting the quantum correlations by interfering the signal and the idler with a beamsplitter, yields measurement results with a higher signal-to-noise ratio, thereby it has a significant advantage over a direct homodyne measurement in the classical scenario. Similarly, the nulling receiver discussed in the main text exploits the quantum correlation with the assistance of an OPA, which yields a significant advantage on a photon number resolving detector. In the following, we calculate the performance of a homodyne detector assisted by OPA, where we find that the advantage remains but is undermined.

Now we consider a homodyne receiver, measuring the quadratures of the signal-idler pair nulled by an OPA with gain similarly defined by Eq. \eqref{eq:G0}. After the OPA, each of the two-mode signal-idler pair is in a Gaussian state with the covariance matrix given by Eq.~\eqref{cov_mat}. For homodyne measurements, one can choose to measure the position $q_S$ or momentum $p_S$ quadrature of the signal mode and $q_I$ or $p_I$ for the idler mode. The two outcomes, defined as $x, y$, are Gaussian random variables characterized by the marginal covariance matrix. 
Observing the symmetry, we see that the measurement on $\{q_S,p_I\}$ has the same statistics with $\{p_S,q_I\}$, ditto for $\{q_S,q_I\}$ with $\{p_S,p_I\}$. Upon obtaining all measurement results across all mode-pairs, we utilize maximum likelihood estimation (MLE) to make the decision. In general, the error rate of this OPA-assisted homodyne receiver has no closed form.

To provide an outline of the practical performance, we give the numerical result of the ideal case $\kappa_B=1$. In this case, the covariance matrix for the background channel is $\Lambda^{(B)}=\bm I$. We find it possible to eliminate the correlation between the two measurement outcomes $\{x,y\}$ per copy by a linear combination that diagonalizes the covariance matrix, mapping $\{x,y\}$ to $\{x', y'\}$. Then we have the weighted M-copy squared sum $s=\sum_{L=1}^{M}ax_{L}'^2+by_{L}'^2$, as a sufficient statistic for the $2M$-dimensional Gaussian, thereby the $2M$-dimensional MLE reduces to 1-dimensional. Here the weights are $a=1/\sigma_{x'}^{2\,(T)}-1$, $b=1/\sigma_{y'}^{2\,(T)}-1$, where the variances $\sigma^2$'s are obtained from the diagonalized covariance matrix and we have used the fact $\sigma_{x'}^{2\,(B)}=\sigma_{y'}^{2\,(B)}=1$ when $\kappa_B=1$. Finally, we can express $s=a\sigma_{x'}^{2\,(h)}X_1+b\sigma_{y'}^{2\,(h)}X_2$ as a weighted sum of two chi-square distributed random variables $X_1, X_2\sim \chi^2 (M)$. The corresponding distribution of $s$ under hypothesis $h$ is a generalized chi-square distribution, which can be numerically calculated. 
Then, the error probability of MLE is numerically obtained by an numerical integration.

\begin{figure}
    \centering
    \includegraphics[width=0.3\textwidth]{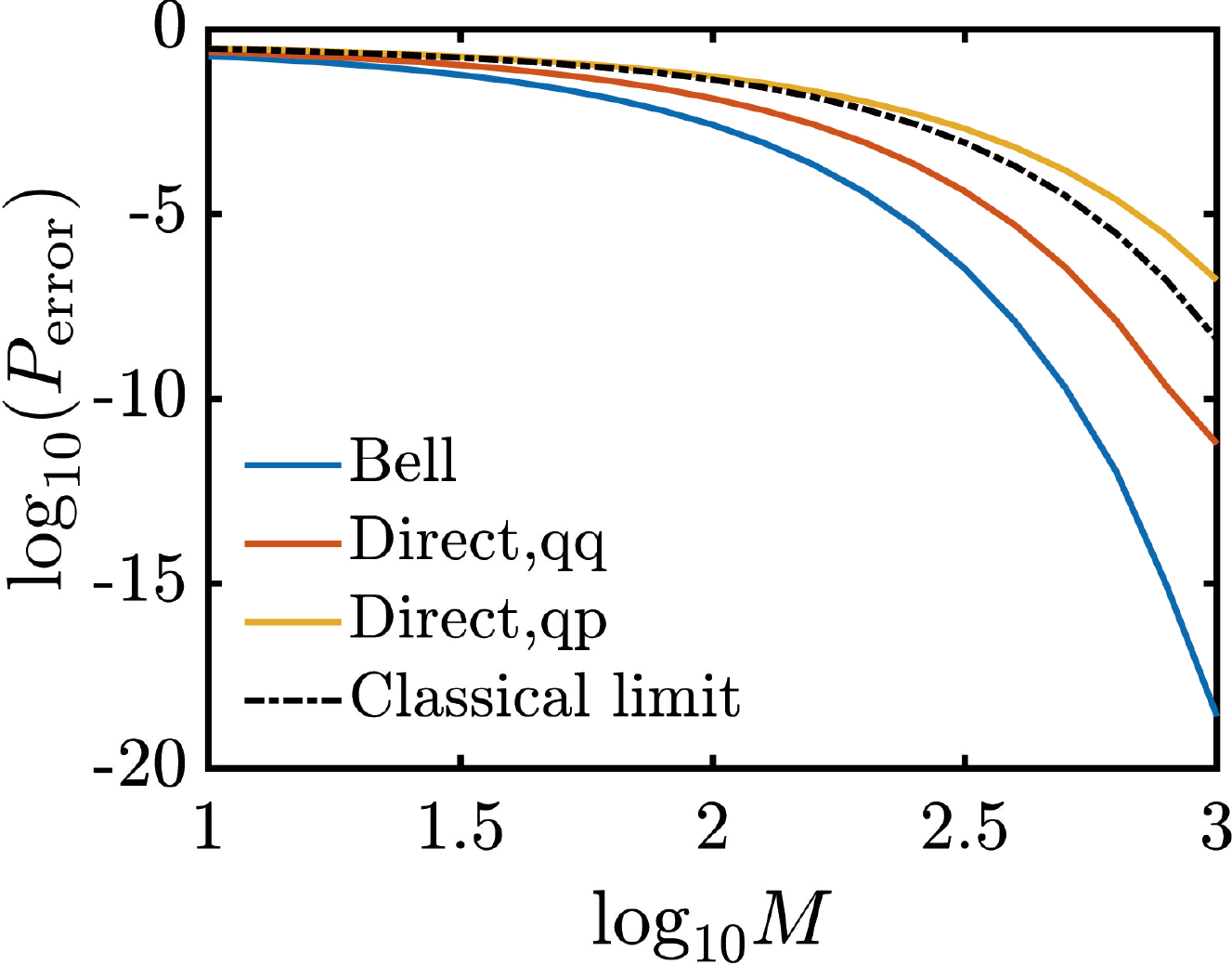}
    \caption{Error rates of the entanglement-assisted (EA) homodyne receivers with $M$ identical copies in the absorption detection scenario. We compare the performance of Bell receiver (blue solid), the OPA-assisted homodyne receivers measuring quadratures $\{q_1,q_2\}$ (red solid) and measuring quadratures $\{q_1,p_2\}$ (orange solid), with the classical limit Eq.~\eqref{C_Helstrom} (black dashed). The mean photon number of the TMSV source is $N_S=1$, the transmissivities of channels are $\kappa_T=0.75$ and $\kappa_B=1$. }
    \label{fig:quadrature}
\end{figure}

Fig.~\ref{fig:quadrature} compares the Bell receiver with the OPA-assisted homodyne receivers and the classical limit, which illustrates the superiority of the Bell measurement. Note that the homodyne receiver measuring $\{q_S,p_I\}$ fails to achieve an advantage over the classical at this moment. This is due to the absence of the quantum correlation between $q_S$, $p_I$. Above all, considering its supremacy among the homodyne receivers, we compare our design only with the Bell receiver in the main text.

\section{Error probability lower bound of classical pattern recognition}
\label{app:general_bound} 
\subsection{Single-mode phase-insensitive Gaussian channels}

As explained in Ref.~\cite{zhuang2020multiple}, the action of a single-mode (covariant) phase-insensitive Gaussian channel over input quadratures $\hat{\bm x}=(\hat{q},\hat{p})^T$ can be represented by the transformation $\hat{\bm x} \rightarrow \sqrt{\mu} \hat{\bm x} + \sqrt{|1-\mu|}\hat{\bm x}_{E} + \xi$, where $\mu$ is a transmissivity ($0 \le \mu \le 1$) or a gain ($\mu \ge 1$), $\hat{\bm x}_{E}$ are the quadratures of an environmental mode in a thermal state with noise variance $\omega = 2N+1$, with $N$ being the mean number of photons, and $\xi$ is additive classical noise, i.e., a random 2-D Gaussian distributed vector with covariance matrix $w_{\rm add}\mathbf{I}$. Here we assume vacuum shot noise equal to $1$. 

Note that, for a coherent state at the input, the output state of the channel is generally thermal with covariance matrix $\mathbf{V} = (\mu + |1-\mu|\omega + \omega_{\rm add})\mathbf{I}$. Setting $\omega = (1+2E-\omega_{\rm add}-\mu)/|1-\mu|$, this matrix simply becomes $(2E+1)\bm I$. Therefore, conditionally on a coherent state input, the channel can be described by the two parameters $\mu$ and $E$, which we denote as $\Phi_{\mu,E}$. In particular, for a thermal-loss channel, we have $1 \le \mu \le 1$, and $E=(\omega-1)(1-\mu)/2=(1-\mu)N$; for a noisy amplifier, we have $\mu \ge 1$, and $E=(\omega+1)(\mu-1)/2=(\mu-1)(N+1)$; and finally, for an additive Gaussian noise channel, we have $\mu = 1$ and $E=\omega_{\rm add}/2$.

\subsection{Ultimate lower bounds for general patterns}
Consider the general case of hypothesis testing between $H$ different patterns of composite channels 
\be 
\calE_n=\otimes_{\ell=1}^m \left(\Phi_{\mu_\ell^{(n)},E_\ell^{(n)}}^{\otimes M}\right)_{S_\ell}, 1\le n \le H,
\label{general_channels}
\ee 
each acting on $m$ subsystems $\{S_\ell\}_{\ell=1}^m$. Here each subsystem consists $M$ modes and each mode goes through a single-mode phase-insensitive bosonic Gaussian channel $\Phi_{\mu_\ell^{(n)},E_\ell^{(n)}}$. Therefore, a global channel $\calE_n$ is specified by coefficients
$\{\mu_\ell^{(n)},E_\ell^{(n)}\}_{\ell=1}^m$. 

\begin{lemma}
\label{lemma:LBgeneral}
Consider classical states (with positive P-representation) as the input, assuming a global energetic constraint of $mMN_S$ mean photons with $M$ modes irradiated over each of the $m$ subsystems $S_k$, the error probability of equal-prior hypothesis testing between $H$ channels $\calE_n$ in Eq.~(\ref{general_channels}) is lower bounded by
\be 
P_H\ge P_{H,LB}^C=\frac{H-1}{2H} \bar{c}^2
\exp\left[-B^\star mMN_S  \right],
\label{general_bound_app}
\ee 
where
\begin{align}
&\bar{c}^K=\prod_{n^\prime >n}\prod_{\ell=1}^m
\nonumber
\\
&\left\{{1+\left[\sqrt{E_\ell^{(n)}(1+E_\ell^{(n^\prime)})}-\sqrt{E_\ell^{(n^\prime)}(1+E_\ell^{(n)})}\right]^2}\right\}^{-M/2} ,
\\
&B^\star=\max_\ell\frac{1}{K}\sum_{n^\prime >n}
\frac{(\sqrt{\mu_\ell^{(n)}}-\sqrt{\mu_\ell^{(n^\prime)}})^2}{1+E_\ell^{(n)}+E_\ell^{(n^\prime)}},
\label{Bstar}
\end{align}
with $K=H(H-1)/2$.

A tighter lower bound 
\be 
P_{H,LB}^C=\frac{K}{H^2}f_\star^2
\label{LB_numerical_app}
\ee 
can be achieved by solving a constrained optimization
\begin{align}
f_\star=\min_{\{X_\ell\}}: &\sum_{n^\prime>n} \frac{1}{K}   C_{n,n^\prime} \exp\left[-\frac{1}{2}\sum_{\ell=1}^mB_\ell^{(n,n^\prime)} X_\ell\right] 
\\
&\mbox{under constraint }\sum_{\ell=1}^m X_\ell \le mMN_S.
\end{align}
The constants
\begin{align}
&B_\ell^{(n,n^\prime)} = \frac{(\sqrt{\mu_\ell^{(n)}}-\sqrt{\mu_\ell^{(n^\prime)}})^2}{1+E_\ell^{(n)}+E_\ell^{(n^\prime)}}
\\
&C_{n,n^\prime}=
\nonumber
\\
&\prod_{\ell=1}^m \left[\frac{1}{1+\big(\sqrt{E_\ell^{(n)}(1+E_\ell^{(n^\prime)})}-\sqrt{E_\ell^{(n^\prime)}(1+E_\ell^{(n)})}\big)^2}\right]^{M/2}.
\end{align}

\end{lemma}
\begin{corollary}
(classical channel-position finding, originally derived in~\cite{zhuang2020multiple})
When $H=m$, 
\be 
\mu_\ell^{(n)}=\left\{ \begin{array}{ll}
         \mu_T & \mbox{if $n =\ell$};\\
        \mu_B & \mbox{if $n \neq \ell$},\end{array} \right.
\ee 
and
\be 
E_\ell^{(n)}=\left\{ \begin{array}{ll}
         E_T & \mbox{if $n =\ell$};\\
        E_B & \mbox{if $n \neq \ell$},\end{array} \right.
\ee 
the pattern recognition problem corresponds to channel-position finding with a target channel $\Phi_{\mu_T,E_T}^{\otimes M}$ among $(m-1)$ background channels $\Phi_{\mu_B,E_B}^{\otimes M}$, where the above bound in Ineq.~(\ref{general_bound_app}) reduces to
\begin{align}
P_{H,LB} = \frac{m-1}{2m} & c_{E_B,E_T}^{2M} \times \nonumber \\ & \exp\left[-\frac{2MN_S(\sqrt{\mu_B}-\sqrt{\mu_T})^2}{1+E_B+E_T} \right], 
\label{LB_Gaussi_app}
\end{align}
with $c_{E_B,E_T}=[1+\big(\sqrt{E_B(1+E_T)}-\sqrt{E_T(1+E_B)}\big)^2]^{-1}$. In particular, for no passive signature ($E_T=E_B \equiv E$), we have the simplification 
\be
P_{H,LB}= \frac{m-1}{2m} \exp\left[-\frac{2MN_S(\sqrt{\mu_B}-\sqrt{\mu_T})^2}{1+2 E} \right],
\label{LB_Gaussi_app2}
\ee
\end{corollary}

\begin{remark}
The corollary is easy to obtain from Lemma~\ref{lemma:LBgeneral}, recognizing $\bar{c}^2=c_{E_B,E_T}^{2M}$ and $\sqrt{\mu_\ell^{(n)}}-\sqrt{\mu_\ell^{(n^\prime)}}$ is zero unless $\ell=n$ or $\ell=n^\prime$, therefore we have
\begin{align}
B^\star=\frac{2}{m} (\sqrt{\mu_B}-\sqrt{\mu_T})^2/(1+E_B+E_T),
\end{align} 
which easily leads to Eq.~(\ref{LB_Gaussi_app}).
\end{remark}

\begin{corollary}
(classical channel-position finding with multiple target channels)
When there are $k$ target channels $\Phi_{\mu_T,E_T}^{\otimes M}$ among $(m-k)$ background channels $\Phi_{\mu_B,E_B}^{\otimes M}$, we have $H=C_m^k$ patterns. For this case, the lower bound in Eq.~(\ref{general_bound_app}) reduces to
\begin{align}
&P_{H,LB}=
\nonumber
\\
&\frac{H-1}{2H} c_{E_B,E_T}^{2Mw_{m,k}}\exp\left[-
\frac{2w_{m,k}MN_S(\sqrt{\mu_B}-\sqrt{\mu_T})^2}{1+E_B+E_T}   \right],
\end{align} 
with 
\be
w_{m,k}\equiv m\frac{C_{m-1}^{k-1}C_{m-1}^k}{C_m^k(C_m^k-1)}=\frac{k C_{m-1}^k}{C_m^k-1}
\ee

For no passive signature ($E_T=E_B \equiv E$) case, the above bound in Ineq.~(\ref{general_bound_app}) reduces to
\be
P_{H,LB}= \frac{H-1}{2H} \exp\left[-\frac{2w_{m,k}MN_S(\sqrt{\mu_B}-\sqrt{\mu_T})^2}{1+2E}   \right].
\label{LB_Gaussi_app2}
\ee
\end{corollary}
\begin{remark}
Similarly, the corollary is easy to obtain from Lemma~\ref{lemma:LBgeneral}. First, one realizes that due to the symmetry, the maximization in $B^\star$ is achieved by an arbitrary $\ell$, e.g. we let $\ell=1$. And then the non-zero contribution to Eq.~(\ref{Bstar}) only comes from patterns with different channels on subsystem $\ell=1$. Slight simplification of Eq.~(\ref{LB_Gaussi_app2}) leads to Eq.~(\ref{LB_Gaussi}) of the main paper.
\end{remark}

\begin{proof}
For the convenience of analysis, we will parameterize a coherent state $\ket{\alpha}$ with the phase and amplitude squared, i.e.,  $\ket{x,\theta}\equiv \ket{\sqrt{x}e^{i\theta}}$, where $x \ge 0$ and $0 \le \theta \le 2\pi$. In this notation, a multi-mode coherent state over the entire system takes the form $\ket{\bm x, \bm \theta}=\otimes_{\ell=1}^m \big(\ket{\bm x_\ell,\bm \theta_\ell}_{S_\ell}\big)$, where each subsystem state $\ket{\bm x_\ell,\bm \theta_\ell}_{S_\ell}=\otimes_{\ell^\prime=1}^M \ket{x_\ell^{(\ell^\prime)},\theta_\ell^{(\ell^\prime)}}$ is again a tensor product of multiple modes with generally-different amplitudes. Here $\bm x_\ell$ are positive and real vectors $\bm x_\ell=(x_\ell^{(1)},\cdots, x_\ell)\equiv \{x_\ell^{(\ell^\prime)}\}_{\ell^\prime=1}^M$ and $\bm x$ is a simple concatenation of them, i.e., $\bm x=(\bm x_1,\cdots, \bm x_m)$.

In this notation, the general classical state
as the input can be written as a Lebesgue integral
\be
\rho=\int{dP}  \ket{\bm x, \bm \theta}\bra{\bm x, \bm \theta},
\ee 
where the probability measure $P$ over $\bm x,\bm \theta$ can be arbitrary. Let us define 
\be
\|\bm x\|_1\equiv\sum_{\ell,\ell^\prime} |x_\ell^{(\ell^\prime)}|=\sum_{\ell,\ell^\prime} x_\ell^{(\ell^\prime)},
\ee
which is the standard one-norm and equals the total mean photon number of the state $\ket{\bm x, \bm \theta}$. Then, the total energy constraint leads to the inequality
\be
\int{dP^\prime} \|\bm x\|_1 \le mMN_S,
\label{energy_constraint_one_norm}
\ee 
where the integral has been simplified to a marginal probability measure $P^\prime$ restricted to the non-negative variables $\bm x$. 

The total conditional state at the output of the channel $\calE_n$ is also a mixture, with expression
\begin{align}
\rho_n^{\rm C}=\calE_n  (\rho)=\int{dP} \rho^{\rm C}_{\bm x, \bm \theta,n}, 1\le n \le H,
\end{align}
where each conditional state is given by
\be
\rho^{\rm C}_{\bm x, \bm \theta,n}=
\otimes_{\ell=1}^m (\rho^{\rm C}_{\mu_\ell^{(n)},E_\ell^{(n)}})_{S_\ell} 
\ee
The state $(\rho^{\rm C}_{\mu_\ell^{(n)},E_\ell^{(n)}})_{S_\ell}$ is a product of $M$ displaced thermal states, each with amplitude $\sqrt{\mu_\ell^{(n)} x_\ell^{(\ell^\prime)}}e^{i\theta_\ell^{(\ell^\prime)}}$ and covariance matrix $(2E_\ell^{(n)}+1)\bm I$.

We use the fidelity-based lower bound of Helstrom limit~\cite{montanaro2008lower},
\begin{equation}
P_H\geq P_{H,LB}\equiv \sum_{k^\prime>k}p_{k^\prime}p_{k}F^{2}(\rho_{k^\prime},\rho_{k}).
\label{montanaro2008lower}%
\end{equation}
From Eq.~(\ref{montanaro2008lower}), we can write the following lower bound to the mean error probability. Consider the equal prior case for simplicity.
\begin{align}
&
P_{H,LB}^{\rm C}=\sum_{n^\prime>n}\frac{1}{H^2} F^2\left[\int{dP} 
\rho_{\bm x,\bm \theta,n}^{\rm C} ,\int{dP}
\rho_{\bm x,\bm \theta,n^\prime}^{\rm C} \right] \nonumber
\\
&
\ge 
\frac{K}{H^2}\sum_{n^\prime>n} \frac{1}{K}
\Big\{\int{dP}   F[
\rho_{\bm x,\bm \theta,n}^{\rm C},
\rho_{\bm x,\bm \theta,n^\prime}^{\rm C}] \Big\}^2 \nonumber
\\
&
\ge 
\frac{K}{H^2}
\Big\{ \sum_{n^\prime>n} \frac{1}{K}\int{dP}   F[\rho_{\bm x,\bm \theta,n}^{\rm C},
\rho_{\bm x,\bm \theta,n^\prime}^{\rm C}]\Big\}^2, \label{toreplace}
\end{align}
where use the joint concavity of fidelity 
\be
F\left[\int{dp_x}  \rho_x ,\int{dp_x}  \sigma_x \right]\ge \int{dp_x}  F[\rho_x,\sigma_x],
\ee 
and Jensen's inequality for the square function with $K=(H-1)H/2$. 

Let us now address each fidelity term 
\begin{align}
&
F^{\rm C}_{n,n^\prime} \equiv F[\rho_{\bm x,\bm \theta, n}^{\rm C},\rho_{\bm x,\bm \theta,n^\prime\neq n}^{\rm C}] \nonumber
\\
&=
F[\otimes_{\ell=1}^m (\rho^{\rm C}_{\mu_\ell^{(n)},E_\ell^{(n)}})_{S_\ell}, \otimes_{\ell=1}^m (\rho^{\rm C}_{\mu_\ell^{(n^\prime)},E_\ell^{(n^\prime)}})_{S_\ell}]
\\
&=\prod_{\ell=1}^m 
F[ (\rho^{\rm C}_{\mu_\ell^{(n)},E_\ell^{(n)}})_{S_\ell},  (\rho^{\rm C}_{\mu_\ell^{(n^\prime)},E_\ell^{(n^\prime)}})_{S_\ell}]
\end{align}
Using Gaussian fidelity formula~\cite{Weedbrook_2012}, we can compute
\begin{align}
&F[ (\rho^{\rm C}_{\mu_\ell^{(n)},E_\ell^{(n)}})_{S_\ell},  (\rho^{\rm C}_{\mu_\ell^{(n^\prime)},E_\ell^{(n^\prime)}})_{S_\ell}]
\nonumber
\\
&=
{\left(c_\ell^{(n,n^\prime)}\right)}^{M/2}\exp\left[-\frac{1}{2}B_\ell^{(n,n^\prime)} \|\bm x_\ell\|_1 \right]
\label{FC_general}
\end{align}
where the constant
\begin{align}
&B_\ell^{(n,n^\prime)} = \frac{(\sqrt{\mu_\ell^{(n)}}-\sqrt{\mu_\ell^{(n^\prime)}})^2}{1+E_\ell^{(n)}+E_\ell^{(n^\prime)}}
\\
&c_\ell^{(n,n^\prime)}=\frac{1}{1+\big(\sqrt{E_\ell^{(n)}(1+E_\ell^{(n^\prime)})}-\sqrt{E_\ell^{(n^\prime)}(1+E_\ell^{(n)})}\big)^2}.
\end{align}
Note that $B_\ell^{(n,n^\prime)}>0$ and $c_\ell^{(n,n^\prime)}\le 1$.
From the one-norm in the expression above, it becomes clear that the performance is exactly the same regardless how the energy is distributed among the $M$ modes impinging on a subsystem, as long as the mean total energy irradiated over the subsystem is fixed. 
Therefore we have
\be 
F^{\rm C}_{n,n^\prime}=C_{n,n^\prime}  \exp\left[-\frac{1}{2}\sum_{\ell=1}^mB_\ell^{(n,n^\prime)} \|\bm x_\ell\|_1 \right],
\ee 
where we defined $C_{n,n^\prime}=\prod_{\ell=1}^m {\left(c_\ell^{(n,n^\prime)}\right)}^{M/2}$. 
By replacing the $F^{\rm C}_{n,n^\prime}$ in Eq.~(\ref{toreplace}), and noticing that $F^{\rm C}_{n,n^\prime}$ does not depend on $\bm \theta$ we find the following lower bound
\begin{align}
P_{H,LB}^{C}
\ge 
\frac{K}{H^2} \Big\{  \int{dP^\prime}    g(\{\|\bm x_\ell\|_1\}_{\ell=1}^m) 
\Big\}^2,
\end{align}
where we define the function 
\begin{align}
&g(\{\|\bm x_\ell\|_1\}_{\ell=1}^m)\equiv 
\nonumber
\\
&\sum_{n^\prime>n} \frac{1}{K}   C_{n,n^\prime} \exp\left[-\frac{1}{2}\sum_{\ell=1}^mB_\ell^{(n,n^\prime)} \|\bm x_\ell\|_1 \right].
\end{align}
We can use the convexity of $e^{-c x}$ (with $c>0$) and Jensen's inequality to move expectation value to the exponent
\begin{align}
P_{H,LB}^{C}
\ge 
\frac{K}{H^2} \Big\{  \sum_{n^\prime>n} \frac{1}{K}   C_{n,n^\prime} \exp\left[-\frac{1}{2}\sum_{\ell=1}^mB_\ell^{(n,n^\prime)} X_\ell\right] 
\Big\}^2,
\end{align}
where $X_\ell=\int dP^\prime \|\bm x_\ell\|_1 $. Note that equality is only achieved when $P^\prime$ corresponds to a delta function. Overall we want to solve the minimization under constraint Ineq.~(\ref{energy_constraint_one_norm})
\begin{align}
f_\star=\min: &\sum_{n^\prime>n} \frac{1}{K}   C_{n,n^\prime} \exp\left[-\frac{1}{2}\sum_{\ell=1}^mB_\ell^{(n,n^\prime)} X_\ell\right] 
\\
&\mbox{under constraint }\sum_{\ell=1}^m X_\ell \le mMN_S.
\end{align}
Then the lower bound would be
\be 
P_{H,LB}^{C}
\ge 
\frac{K}{H^2}f_\star^2.
\ee 
This gives the lower bound in Eq.~(\ref{LB_numerical_app}).

Below we obtain a further lower bound.
\begin{align}
&g(\{\|\bm x_\ell\|_1\}_{\ell=1}^m)
\nonumber
\\
&\ge 
\left\{
\prod_{n^\prime >n}
\left(
C_{n,n^\prime} \exp\left[-\frac{1}{2}\sum_{\ell=1}^mB_\ell^{(n,n^\prime)} \|\bm x_\ell\|_1 \right]
\right)
\right\}^{1/K}
\\
&=
\bar{c}
\exp\left[-\frac{1}{2K}\sum_{n^\prime >n}\sum_{\ell=1}^mB_\ell^{(n,n^\prime)} \|\bm x_\ell\|_1 \right]
\end{align}
where we have introduced $\bar{c}=\left\{
\prod_{n^\prime >n}C_{n,n^\prime} \right\}^{1/K}$. We have also used $K=H(H-1)/2$ and the fact that arithmetic mean is greater than geometric mean. The equality holds if and only if $c_{n,n^\prime} \exp\left[-\sum_{\ell=1}^mB_\ell^{(n,n^\prime)} \|\bm x_\ell\|_1 \right]$ is equal for all $n,n^\prime$.

Thus overall we may write 
\begin{align}
&
P_{H,LB}^{C}
\ge
\frac{\bar{c}^2 K}{H^2}
\Big\{  \int{dP^\prime}  \exp\left[-\frac{1}{2K}\sum_{n^\prime >n}\sum_{\ell=1}^mB_\ell^{(n,n^\prime)} \|\bm x_\ell\|_1 \right] \Big\}^2 \nonumber \\
&\ge
\frac{\bar{c}^2 K}{H^2}
\Big\{\exp\left[-\int{dP^\prime}\frac{1}{2K}\sum_{n^\prime >n}\sum_{\ell=1}^mB_\ell^{(n,n^\prime)} \|\bm x_\ell\|_1 \right]\Big\}^2 \nonumber
\\
&=
\frac{H-1}{2H} \bar{c}^2
\exp\left[-\frac{1}{K}\sum_{n^\prime >n}\sum_{\ell=1}^mB_\ell^{(n,n^\prime)} X_\ell \right].
\label{ineq_lb_close}
\end{align}
For the second inequality, we use the convexity of $e^{-c x}$ (with $c>0$) and Jensen's inequality to move expectation value to the exponent. 

To minimize the lower bound in Eq.~(\ref{ineq_lb_close}), we need to solve the constrained (by constraint in Eq.~(\ref{energy_constraint_one_norm})) optimization
\begin{align}
&\max \sum_{\ell=1}^m \frac{1}{K}\sum_{n^\prime >n}B_\ell^{(n,n^\prime)} X_\ell
\nonumber
\\
&\mbox{under constraint } \sum_{\ell=1}^m X_\ell \le mMN_S.
\end{align}
The solution to the maximization is simple, denote $\ell^\star=\arg\max_\ell \sum_{n^\prime >n}B_\ell^{(n,n^\prime)}$ and $B^\star=\max_\ell\frac{1}{K}\sum_{n^\prime >n}B_\ell^{(n,n^\prime)}$, and let
\be
X_{\ell^\star}=mMN_S, X_{\ell\neq \ell^\star}=0,
\ee 
which leads to the maximum $B^\star mMN_S$.

Overall we have the lower bound
\be 
P_{H,LB}^C\ge \frac{H-1}{2H} \bar{c}^2
\exp\left[-B^\star mMN_S  \right],
\ee 
It is easy to check that the lower bound can be reached only if
\begin{align}
c_{n,n^\prime} \exp\left[-B_{\ell^\star}^{(n,n^\prime)} mMN_S \right]=\mbox{const}, 
\end{align}
is independent of $n,n^\prime$. This is not always possible and therefore the lower bound is only achievable in certain symmetric cases. In symmetric cases, when $\max_\ell\frac{1}{K}\sum_{n^\prime >n}B_\ell^{(n,n^\prime)}$ are equal for all $\ell$, then one can evenly distribute the energy to achieve the lower bound.

\end{proof}

\begin{figure}
    \centering
    \includegraphics[width=0.45\textwidth]{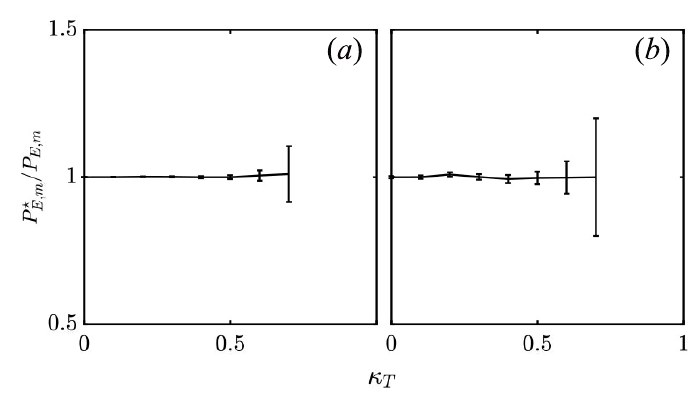}
\caption{Verification of the analytical formulae. We compare the numerical result $P_{E,m}^{\star}$ with the analytical result $P_{E,m}$. $M$ is valued such that the classical lower bound $P_{c,m,LB}=10^{-1}$. (a) the absorption detection Eq.~(\ref{eq:analyt_app}). $\kappa_B=1, N_S=1$.  (b) Verification of the peak positioning Eq.~(\ref{eq:PPManalyt_app}). $\kappa_B=1, N_S=1, m=100$. In the simulation we take $10^8$ samples for the absorption detection case and $2\times 10^6$ samples for the peak positioning case.
 \label{fig:analytical}
}
\end{figure}

\begin{figure}
    \centering
    \includegraphics[width=0.3\textwidth]{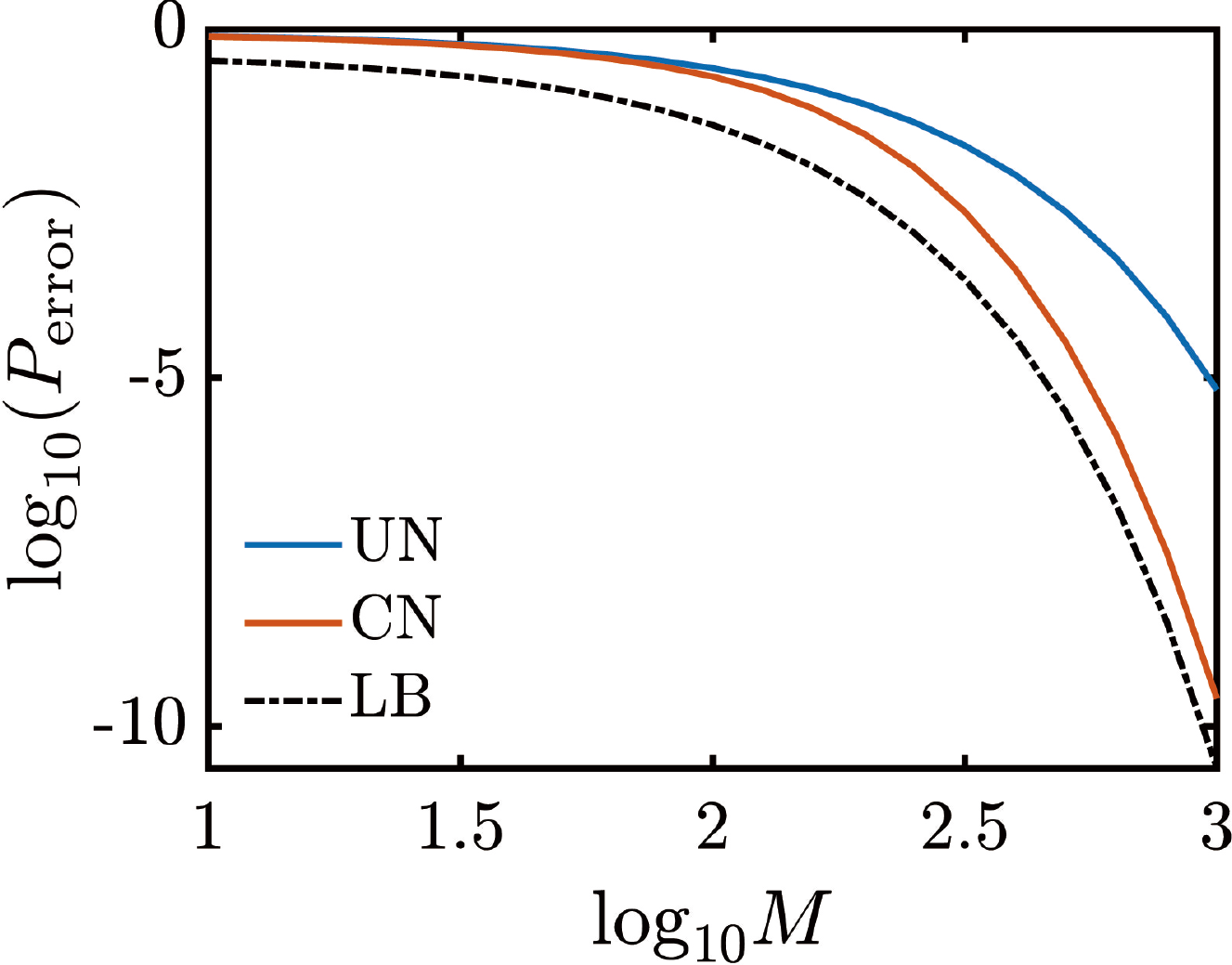}
    \caption{Error rate of the classical nuller with $M$ identical copies in the 100-slot peak positioning scenario. We compare the unconditional nuller (UN, blue solid) and the conditional nuller (CN, red solid) with the classical lower bound Eq.~\eqref{LB_Gaussi} in the main text (LB, black dashed). Source mean photon number $N_S=1$, channel transmissivity $\kappa_T=0.75,\,\kappa_B=0.95$.
    \label{fig:CNvsUNvsLB}
    }
\end{figure}

\begin{figure*}
    \centering
    \subfigure{
    \includegraphics[width=1\textwidth]{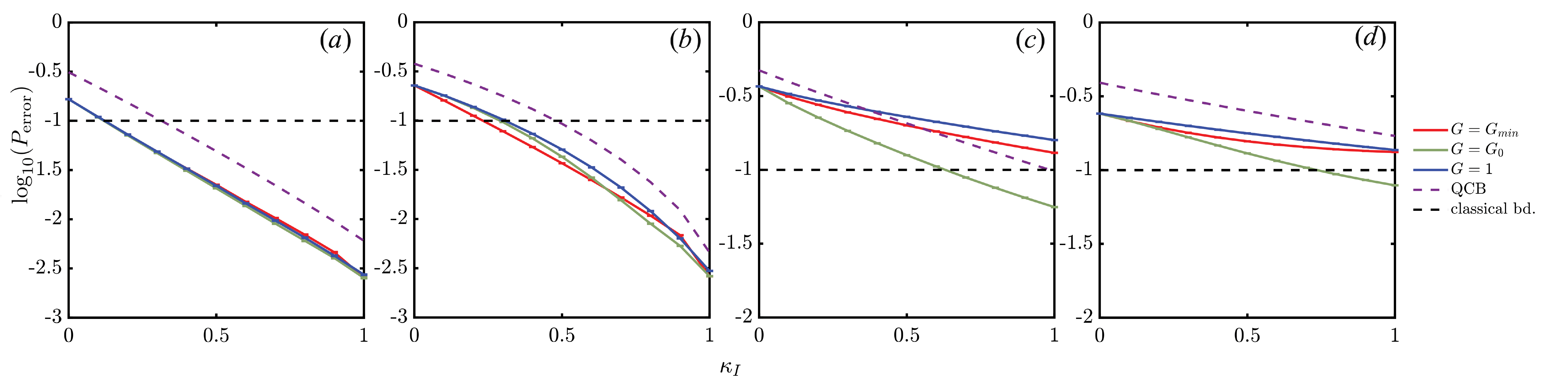}
    }
    \centering
    \subfigure{
    \includegraphics[width=1\textwidth]{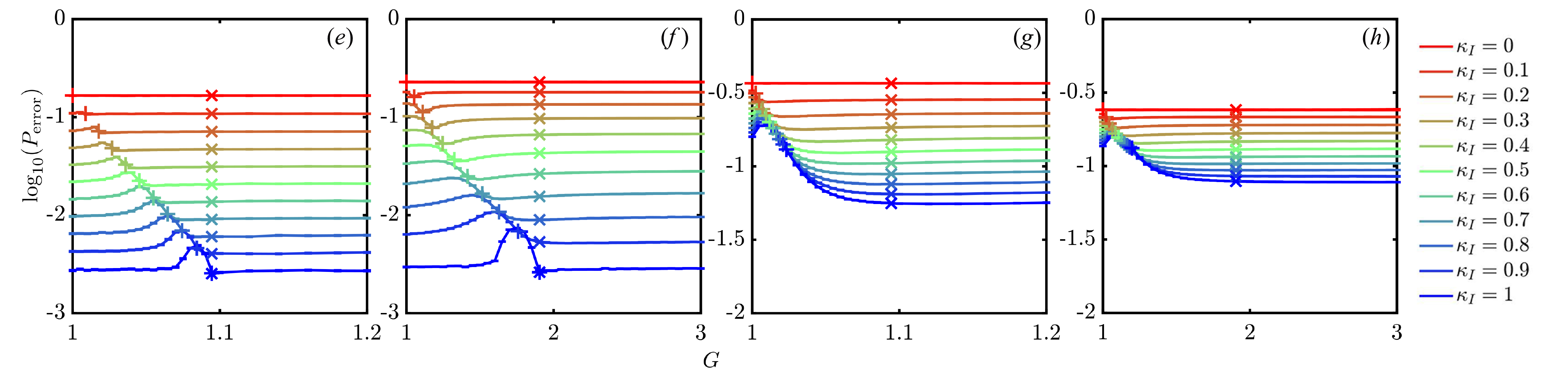}
    }
    \caption{The effect of idler storage efficiency $\kappa_I$ (a-d) and the anti-squeezing gain $G$ (e-h) on the performance of our entanglement-assisted nulling receiver. (a)(e)$N_S=0.1,\,N_B=0$. (b)(f)$N_S=1,\,N_B=0$. (c)(g)$N_S=0.1,\,N_B=1$; (d)(h)$N_S=1,\,N_B=1$. For the gain $G$, we compare the consequential error rate between a wide range of values, including $G_0$, the default setting in the main text, and $G_{min}$, which minimizes the signal mode thermal fluctuation. With the parameter setting in this figure $G_{min}<G_0$. Identical mode number $M$ is chosen such that the classical lower bound $P_{C,1,LB}$ (black dashed line) is fixed to be $0.1$. $\kappa_T=0.75, \kappa_B=0.95$. \QZ{Note that in (c) the QCB and classical bound coincidentally get close when $\kappa_I=1$, this is due to finite $M$ and they will deviate as $M$ further increases.} 
            \label{fig:optimalG}
    }
\end{figure*}

\section{Classical strategy: nulling receiver for coherent state input}

Refs.~\cite{dolinar_processing_1973,dolinar1982near} have conducted a comprehensive analysis on the classical nulling receiver in the noiseless scenario $N_B=0$. We briefly summarize the pertinent conclusions here, and compare the performance with our lower bound. The channel setup in the classical scenario is almost identical to that in the quantum scenario, the only difference being that the source is classical, e.g. coherent states generated by a laser.

\subsection{Absorption Detection} In this scenario, the task is to distinguish the target channel with transmissivity $\kappa_T$ with the background channel $\kappa_B$. Given the source state being $M$ copies of coherent state $\ket{\sqrt{N_S}}$ with total mean photon number $MN_S$, the optimal error rate is tightly bounded by the Helstrom limit~\cite{Qreading}  
\be
P_H=\frac{1}{2}[1-\sqrt{1-e^{-(\sqrt{\kappa_T}-\sqrt{\kappa_B})^2MN_S}}]
\,.\ee
Dolinar has proposed an adaptive receiver \cite{dolinar_processing_1973} that reaches the Helstrom bound. In this case, the improved classical ultimate lower bound Eq.~(\ref{C_Helstrom}) in the main text coincides with this bound, which results from the fact that encoding with two pure states instead of an ensembles, i.e. concentrating the source energy on a single amplitude level, has achieved the optimum.

\subsection{Peak Positioning}
Consider single-peak position among $m$ slots. We categorize the nulling receiver into unconditional nuller and condition nuller. The unconditional nuller, which motivated our entanglement-assisted design in the main text, applies identical displacement $D(\sqrt{\kappa_BN_S})$  to all $m$ modes. In the noiseless case $N_B=0$, the return is in coherent states. The uniform displacement nulls the returned copies through the background channels to vacuum and those through the target channel to yet a coherent state with displacement $\sqrt{\kappa_TN_S}-\sqrt{\kappa_BN_S}$. Then we apply photon counting on every mode, which immediately identifies the target channel if any click is detected. The error only occurs when the coherent state associated with the target channel yields zero photon count with the false-negative error rate $p=\exp[-(\sqrt{\kappa_T}-\sqrt{\kappa_B})^2MN_S]$ for source mean photon energy $MN_S$. The error rate of unconditional nuller is then 
\be 
P_{C,UN}=\frac{m-1}{m}p= \frac{m-1}{m}\exp[-MN_S(\sqrt{\kappa_B}-\sqrt{\kappa_T})^2].
\ee

The conditional nuller, however, applies a mode-by-mode sequence of displacements dependent on the prior measurement results. Specifically, we null the first mode and measure its photon count. If no photon is detected at the first mode, our hypothesis that the target channel be at the first mode is partially confirmed, and we forgo the nulling on the remaining modes unless any photon is detected in the subsequent measurements. Note that in this noiseless case the false positive error rate is zero, any nonzero photon count is a conclusive evidence in favor of rejecting the current hypothesis for the currently measured mode, which immediately gives the conclusion if the rejected hypothesis is `background'. On the other hand, if any photon is detected at the first mode, the target hypothesis is conclusively rejected, we move forward to the next hypothesis that the target channel be at the second mode. In sum, the error only occurs if both measurements on the target channel and the background channel mistook by target yield false negative errors. By iteration the relation $P_{C,m}=[(1-p)P_{C,m-1}+p^2]\times (m-1)/m$, the error rate of the conditional nuller is 
\be
P_{C,CN}=\frac{(1-p)^m+mp-1}{m}
\,.\ee
When $M\gg1$, we have 
\be 
P_{C,CN}\sim \frac{m-1}{2}{\exp}[-2MN_S(\sqrt{\kappa_B}-\sqrt{\kappa_T})^2].
\ee 
In this limit, the conditional nuller loses to the classical lower bound Eq.~\eqref{LB_Gaussi_app2} (here $E=0$) merely by a constant factor $m$, achieving the bound in the exponent indeed.

Fig. \ref{fig:CNvsUNvsLB} compares the two nulling receivers with the classical lower bound Eq.~\eqref{LB_Gaussi} in the main text. It is verified that the unconditional nuller is overwhelmed by the conditional nuller. Furthermore, the latter is shown close to the classical lower bound in the decaying rate, which never achieves the bound though as expected.

\section{Analytical solutions of the error probability}
\label{analytical_cases}
Here we present analytical solutions for the error probability in absorption detection and single-peak positioning, when $\kappa_B=\kappa_I=1, N_B=0$.

\subsection{Absorption detection}

In this case, Eq.~(\ref{eq:G0}) yields the gain $G=1+N_S$. At this moment the squeezing $\calS$ is exactly the inverse of the two-mode squeezing operation that creates the TMSV state from vacuum. From the covariance matrix ${\mathbf{{\mathbf{\Lambda}}}}\left({\kappa_B}\right)$ in Eq.~(\ref{cov_mat_kB}), the background signal $\Xi^{(B)}$ is nulled to the vacuum state, a pure state capable to be discriminated with zero error rate. Applied on the photon statistics $P_m(\bm n_S,\bm n_I|h)$, the maximum-likelihood decision rule accepts the hypothesis $\tilde h=T$ for all the photon count results except for zero counts, which leads to the hypothesis to $\tilde h=B$. In this case, the error only happens when the nulled target-present state $\calS(\Xi^{(T)})$ yields no photon count. Hence the error rate is $P_{E,1}=P_m(\bm 0,\bm 0|T)/2$. Given the transmissivities of background and target present channels $\kappa_B$, $\kappa_T$, and the mean photon number constraint on the source $N_S$, we have the covariances ${\mathbf{{\mathbf{\Lambda}}}}\left({\kappa_T}\right)$ of each pair in $\calS(\Xi^{(T)})$ given by Eq.~(\ref{cov_mat}) with 
\begin{align}
E&=1 + 2N_S(1 + N_S)(1 - \sqrt{\kappa_T})^2,\\
S&=1 + 2 N_S [(1-\sqrt{\kappa_T})(2 + N_S (1- \sqrt{\kappa_T}))],\\
C&=2[(N_S(1-\sqrt{\kappa_T})+1)(1-\sqrt{\kappa_T})\sqrt{N_S(1 + N_S)}].
\end{align}
Thus 
\begin{align} 
P_{E,1}&=P(\bm 0,\bm 0|T)/2=[P(0,0)]^M/2
\\
&=\frac{(C^2+E+S-ES-1)}{C^2},
\end{align}
where $P(n_S,n_I)$ is given by Eq.~(\ref{eq:prob}). Finally
\be
P_{E,1}=\frac{1}{2}\left[\frac{1}{1+N_S(1-\sqrt{\kappa_T})}\right]^{2M}.
\label{eq:analyt_app}
\ee

\begin{figure*}
    \centering
    \subfigure[$N_S=0.1$. $R_E/R_{QCB}$]{
    \includegraphics[width=0.23\textwidth]{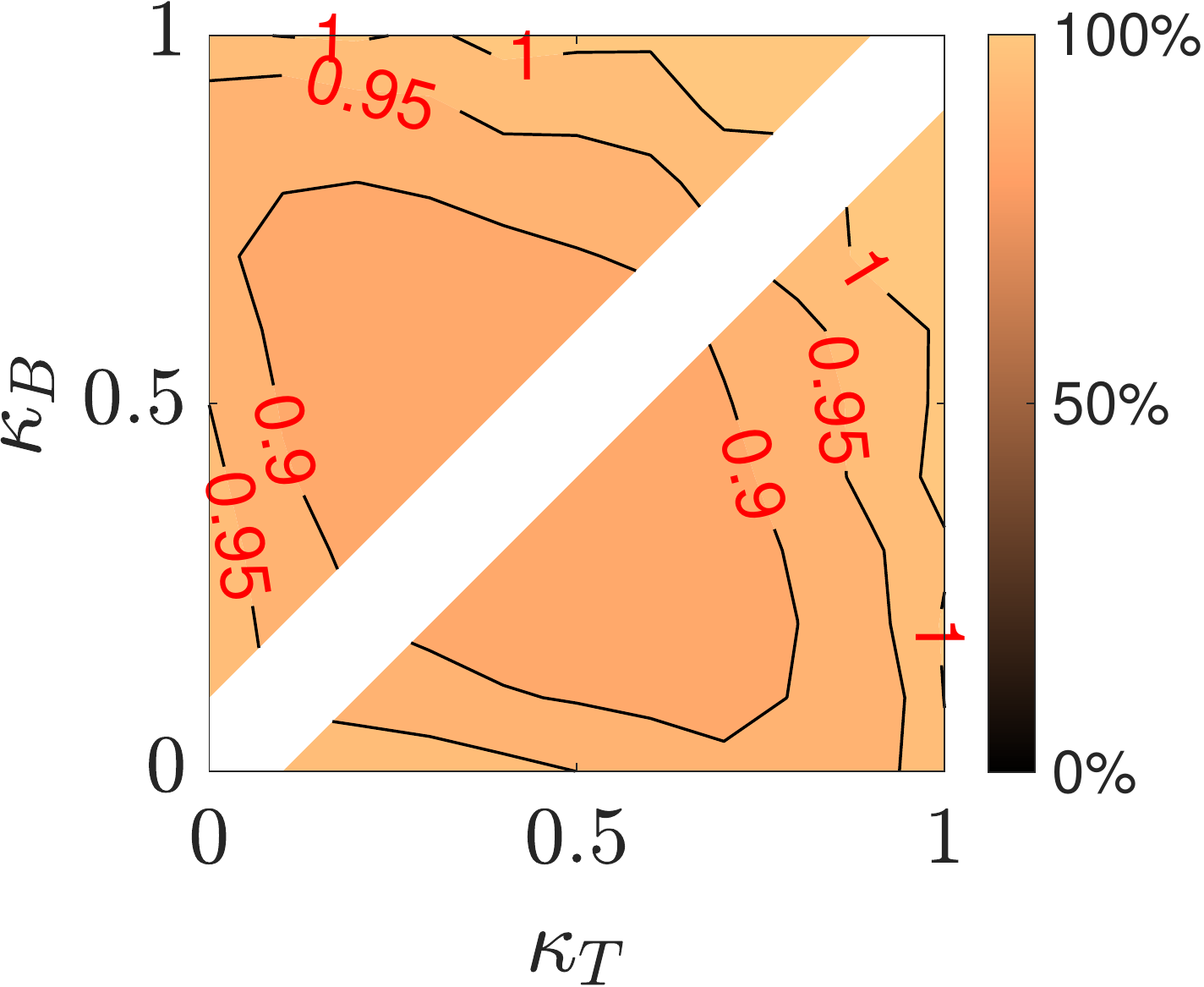}
    }
    \subfigure[$N_S=1$. $R_E/R_{QCB}$]{
    \includegraphics[width=0.23\textwidth]{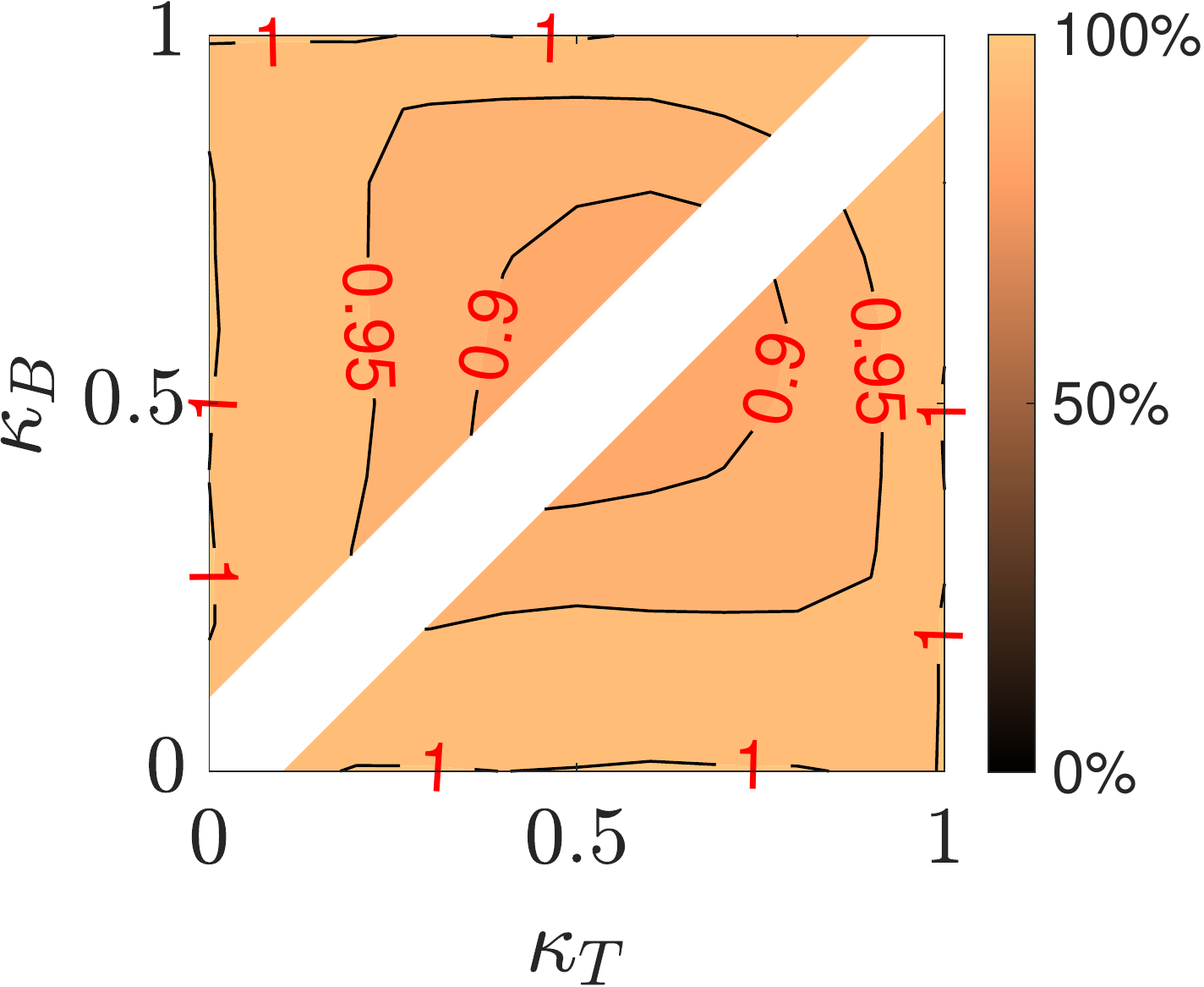}
    }
    \subfigure[$N_S=0.1$. $R_E/E_C$]{
    \includegraphics[width=0.23\textwidth]{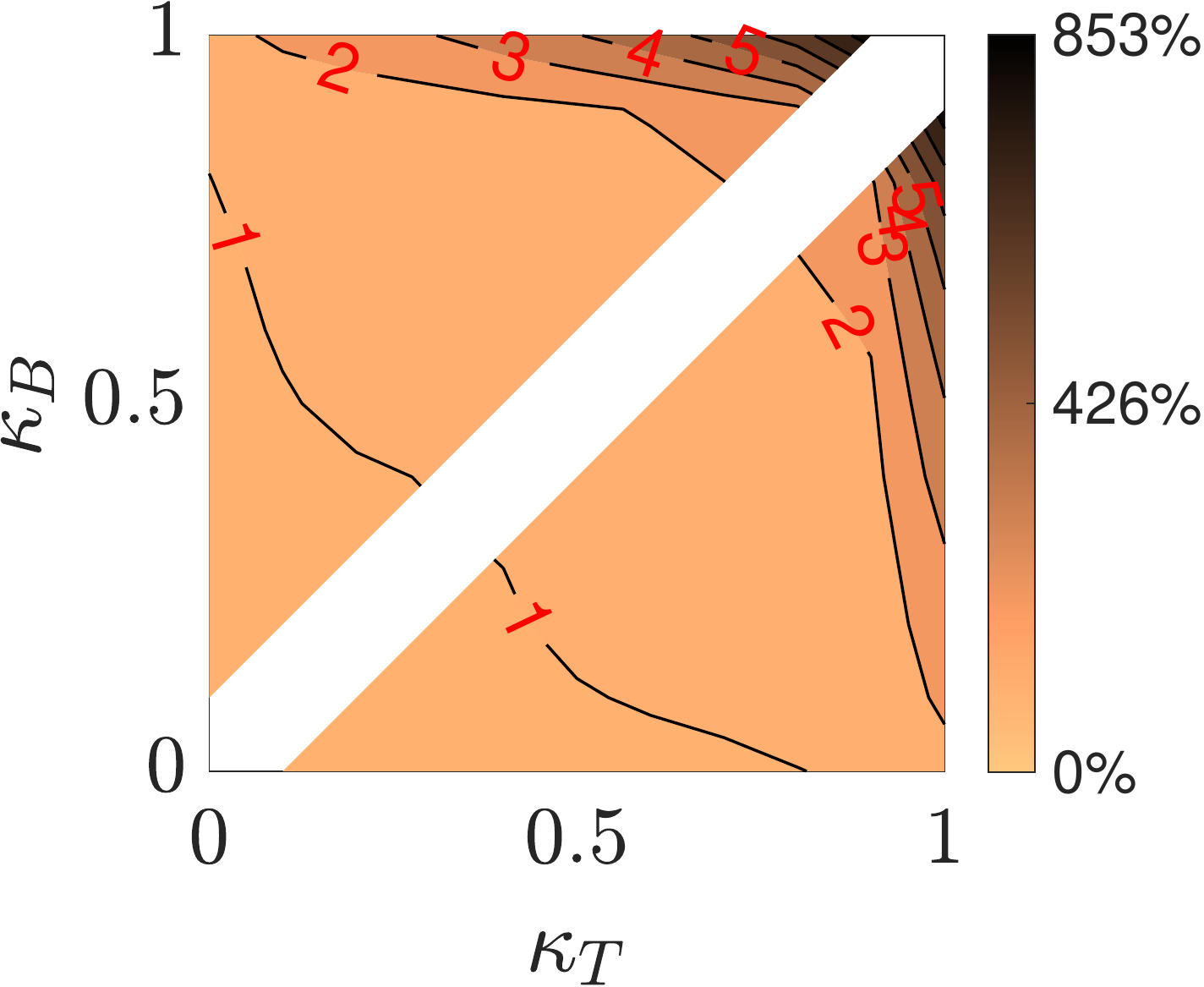}
    }
    \subfigure[$N_S=1$. $R_E/E_C$]{
    \includegraphics[width=0.23\textwidth]{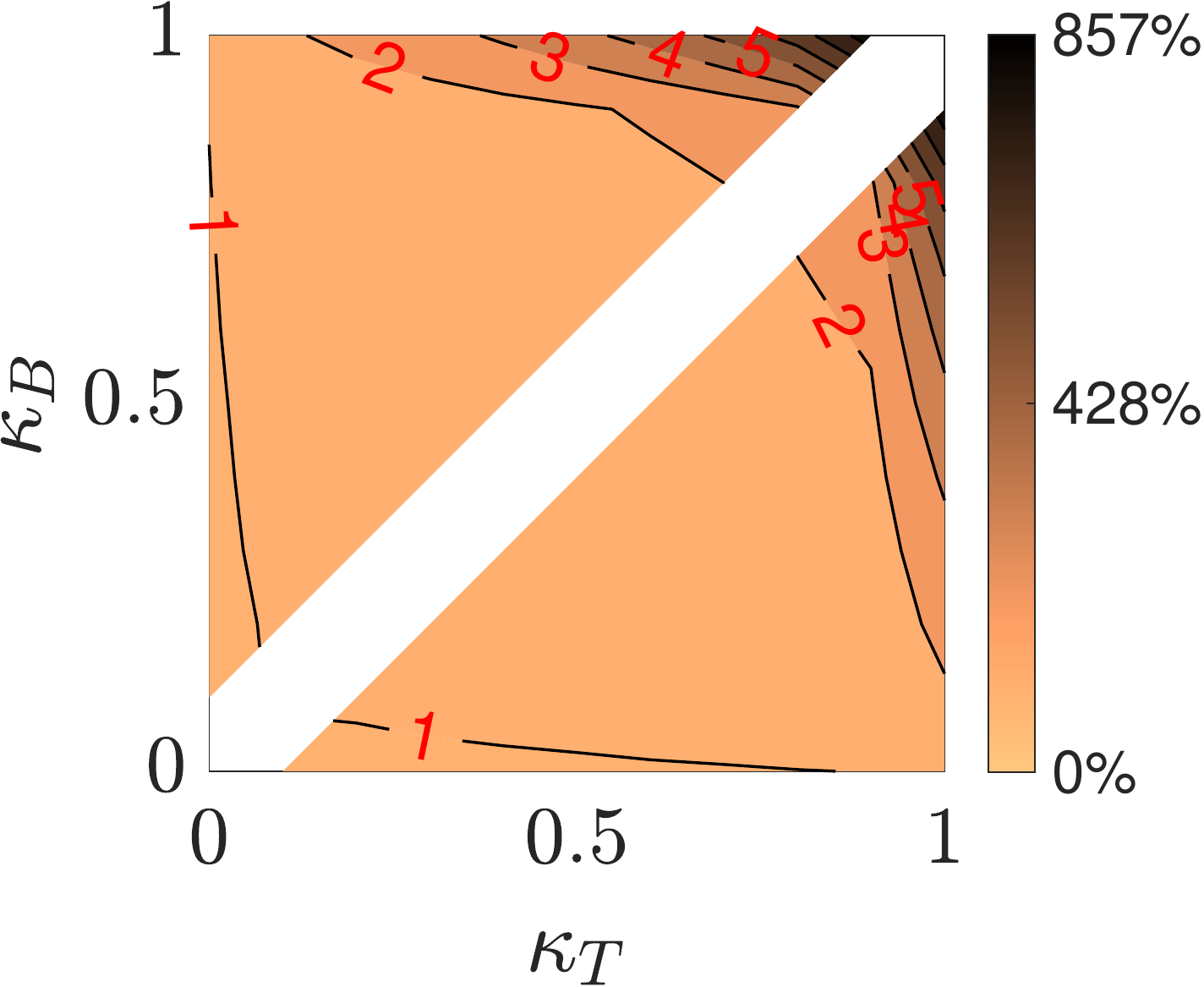}
    }
     \caption{(a)(b) Error exponent ratio $R_E/R_{QCB}$ of noiseless absorption detection over QCB with respect to the transmissivity $\kappa_B, \kappa_T$. Based on Monte Carlo simulation with $5\times10^6$ ($N_S=0.1$) and $5\times10^7$ ($N_S=1$) samples. Standard deviation is (a) below $1\%$ in all data points. (b) below $0.3\%$ in all data points. The difference in the standard deviation results from the fact that more modes are required in (a), therefore the Monte Carlo simulation is more costly and the sample size is traded off. (c)(d) Error exponent ratio of $R_E/E_C$ in the same parameter setting. 
     \label{fig:quReadvsQCB-5}
     }
\end{figure*}

\begin{figure}[!h]
    \centering
    \includegraphics[width=0.4\textwidth]{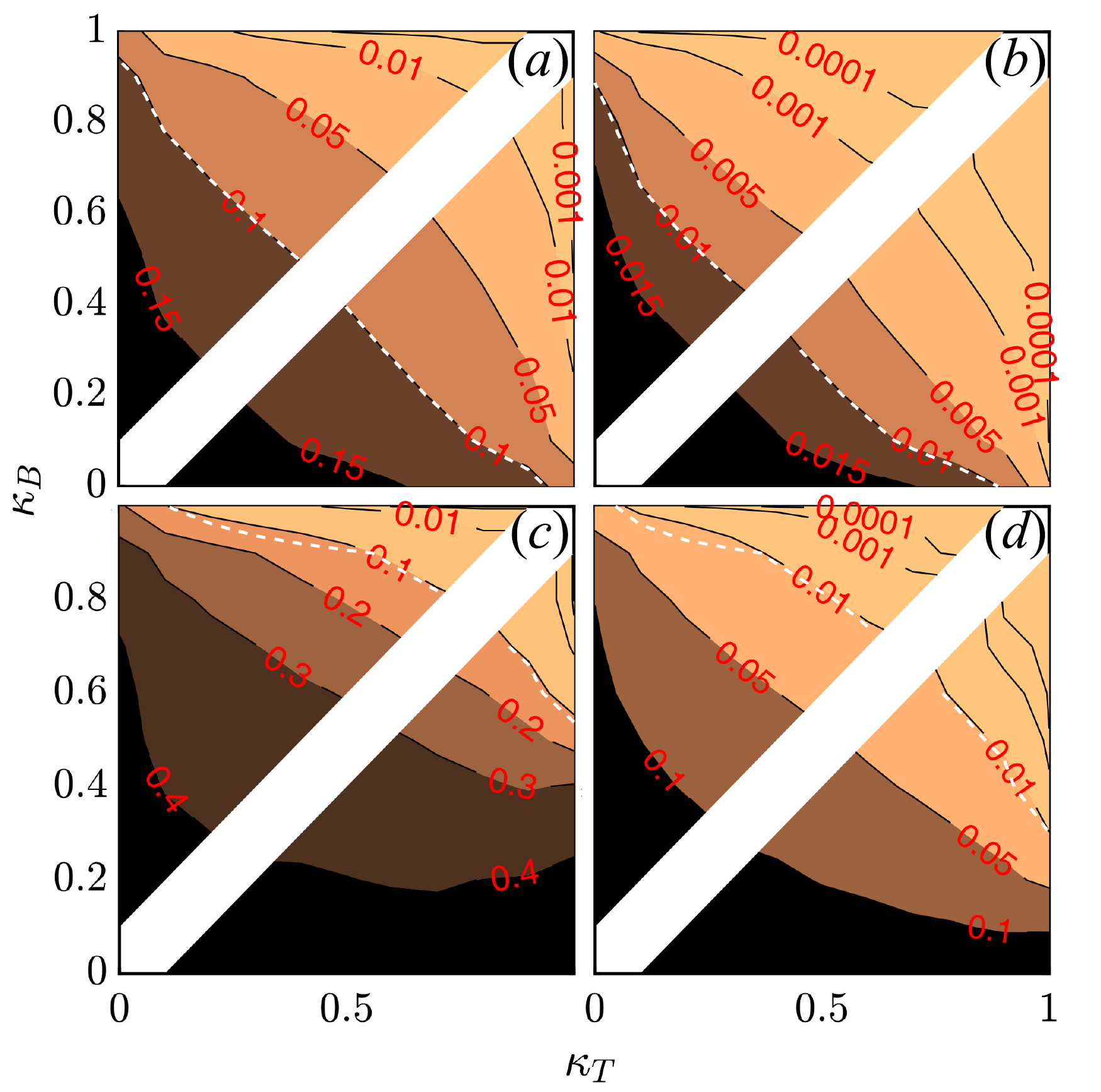}
     \caption{
     Error probability of EAAS versus transmissivities of the background and target channels, $\kappa_{B}$ and $\kappa_{T}$, with different $P_{C,m,LB}$ fixed. Source mean photon number $N_S=0.1$. The white dashed line corresponds to the value of the classical benchmark $P_{C,m,LB}$, therefore indicating the boundary between regions with/without quantum advantage. (a-b) We consider absorption detection case ($m=1$), with $M$ such that~$P_{C,1,LB}=0.1,0.01$. (c-d) We consider single-peak positioning with $m=100$ and $M$ such that~$P_{C,100,LB}=0.1,0.01$.  Monte Carlo simulation sample size $10^5$.
     \label{fig:contours_ideal}
     }
    \centering
    \includegraphics[width=0.4\textwidth]{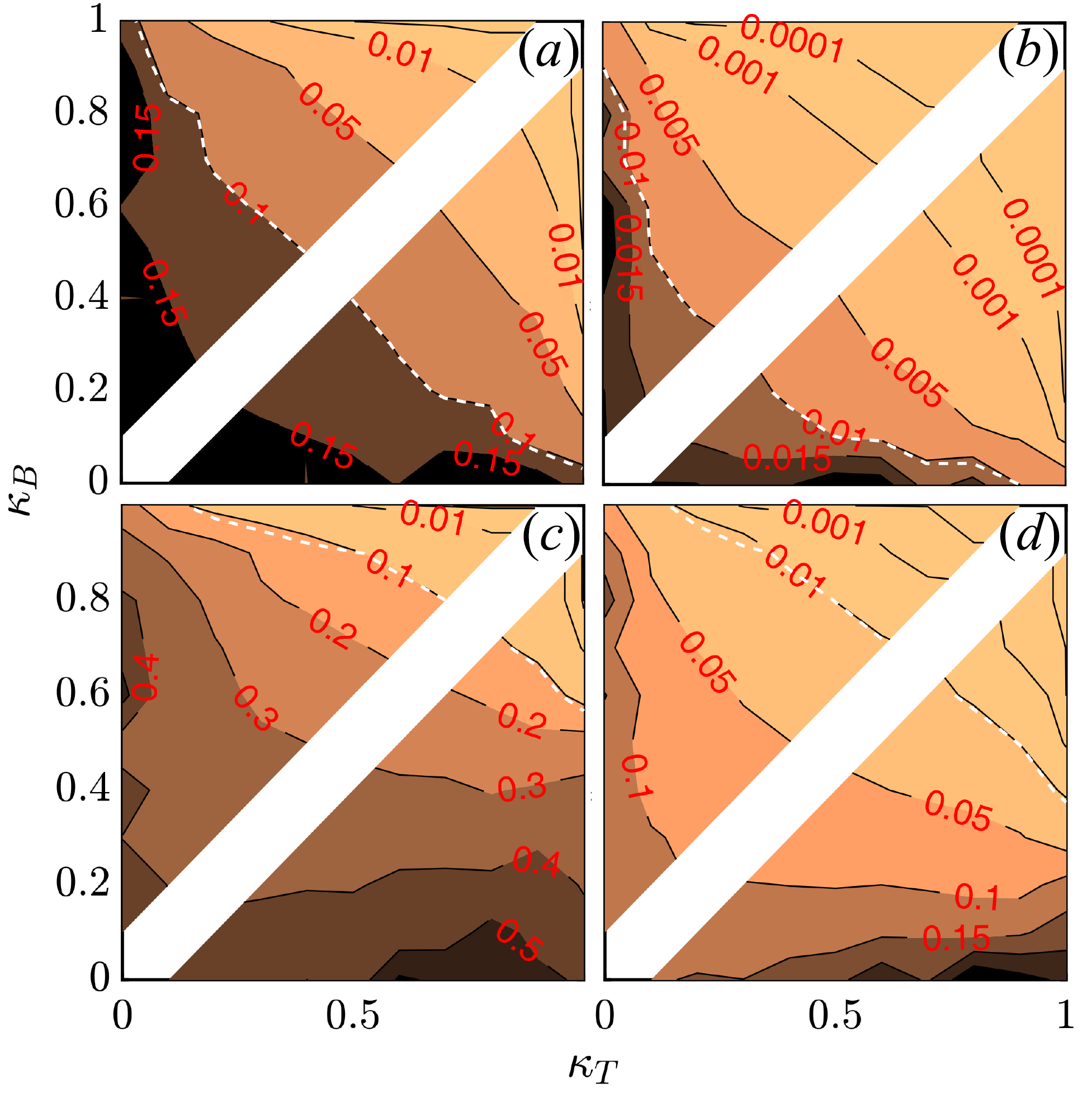}
     \caption{
     Error probability of EAAS versus transmissivities of the background and target channels, $\kappa_{B}$ and $\kappa_{T}$, with different $P_{C,m,LB}$ fixed. Source mean photon number $N_S=1$. The layout is the same as Fig.~\ref{fig:contours_ideal}.  Monte Carlo simulation sample size $10^6$.
     \label{fig:contours_ideal2}
     }
     \end{figure}
     \,
\begin{figure}[!h]
    \centering
    \includegraphics[width=0.4\textwidth]{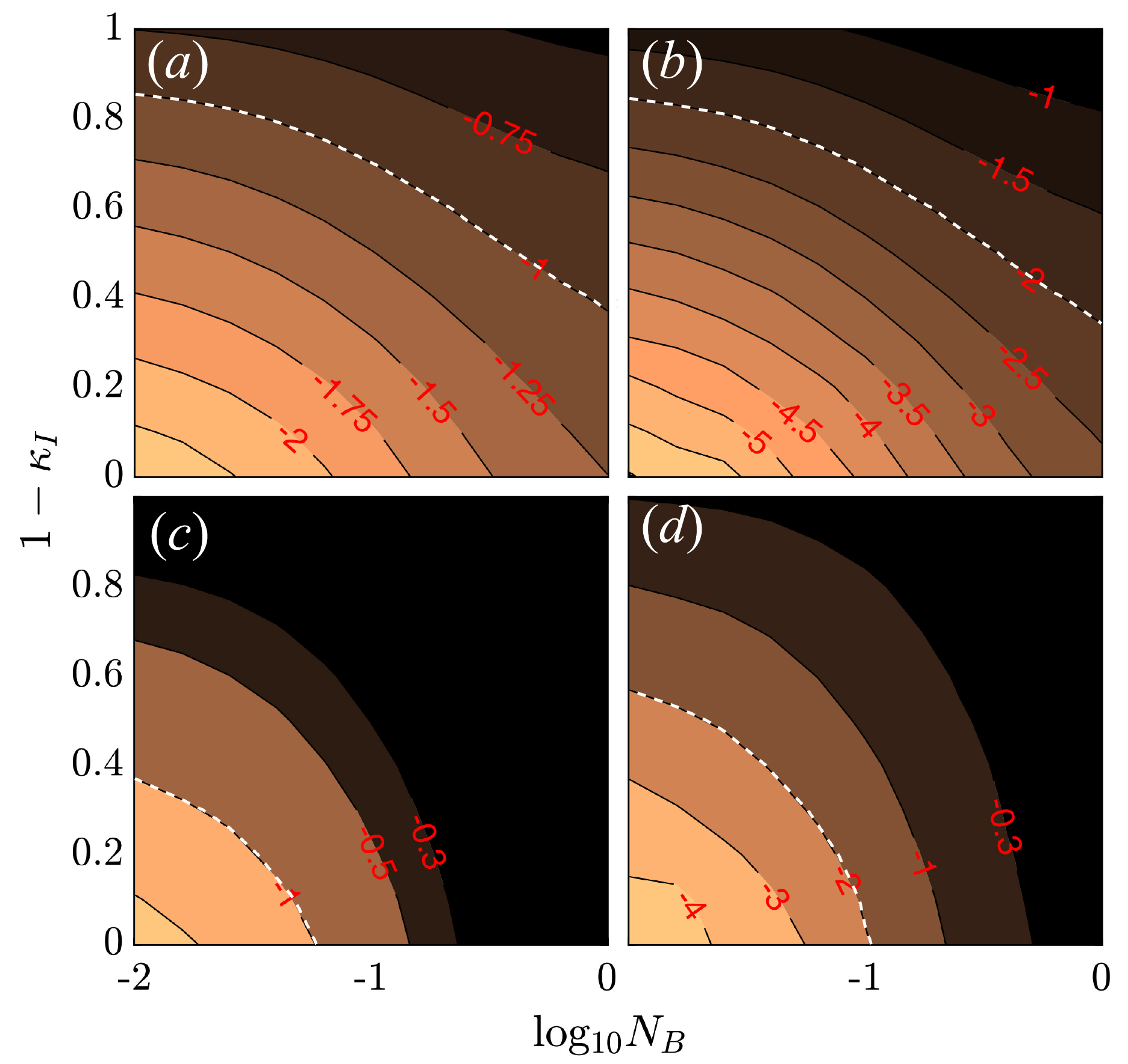}
    \caption{Logarithmic error rate $\log_{10}\left(P_{E,m}\right)$ of EAAS with respect to the thermal noise $N_B$ and the idler loss $1-\kappa_I$ with different $P_{C,m,LB}$ fixed. Parameters are: $N_S=0.1$, $\kappa_T=0.75$, and $\kappa_B=0.95$. The white dashed line corresponds to the value of the classical benchmark $P_{C,m,LB}$, therefore indicating the boundary between regions with/without quantum advantage. (a-b) We consider absorption detection case ($m=1$), with $M$ such that~$P_{C,1,LB}=0.1,0.01$. (c-d) We consider single-peak positioning with $m=100$ and $M$ such that~$P_{C,100,LB}=0.1,0.01$. Monte Carlo simulation sample size $10^5$.
    \label{fig:kINbcontour}
    }
    \centering
    \includegraphics[width=0.4\textwidth]{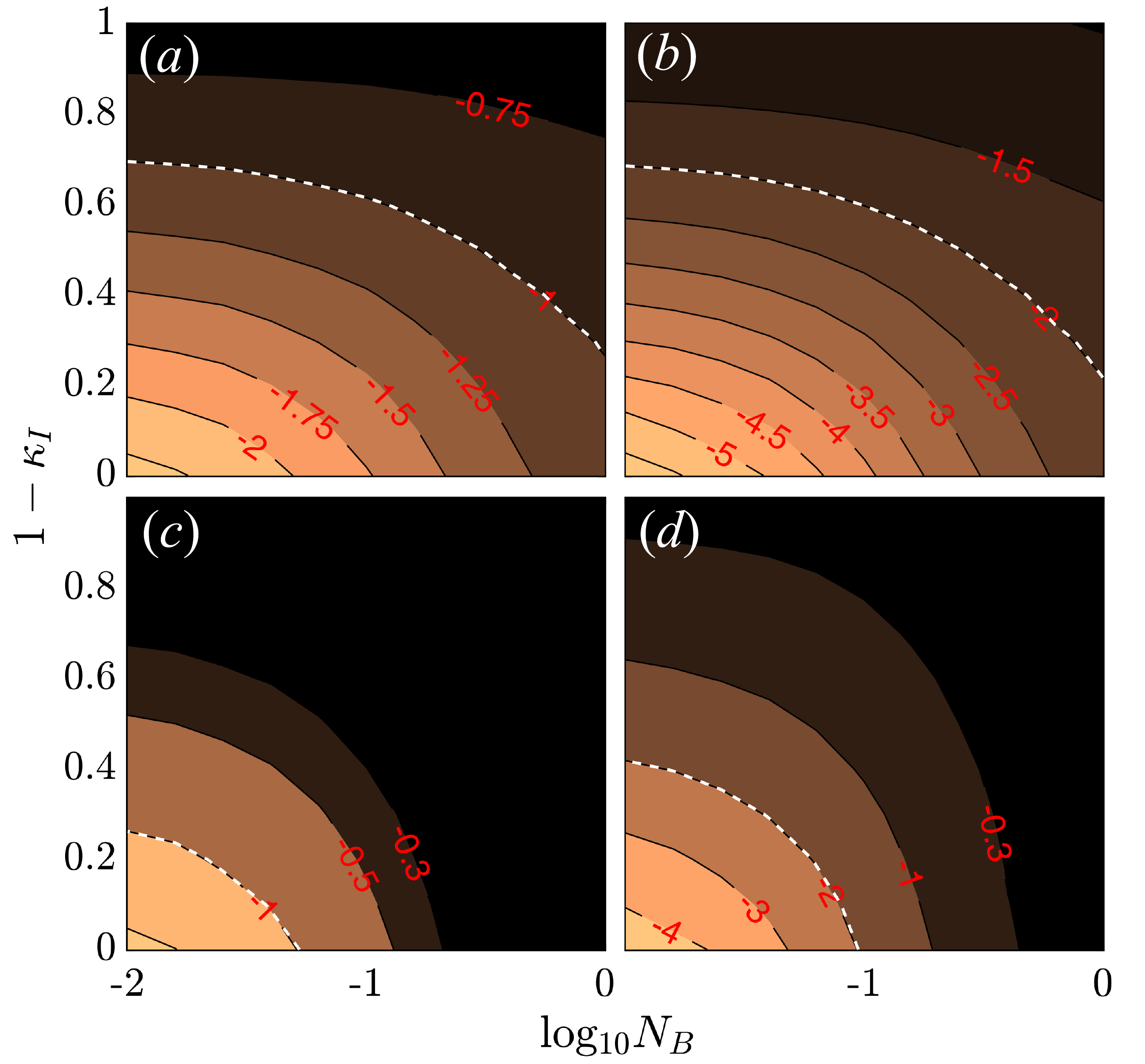}
    \caption{Logarithmic error rate $\log_{10}\left(P_{E,m}\right)$ of EAAS with respect to the thermal noise $N_B$ and the idler loss $1-\kappa_I$ with different $P_{C,m,LB}$ fixed. Parameters are: $N_S=1$, $\kappa_T=0.75$, and $\kappa_B=0.95$. The layout is the same as Fig.~\ref{fig:kINbcontour}. Monte Carlo simulation sample size $10^6$.
    \label{fig:kINbcontour2}
    }
\end{figure}

\subsection{Peak positioning}
The simplest case is when $\kappa_B=1$, which is analytically solvable. First, the receiver nulls all the returned slot by uniformly applying the squeezing process $\calS$ with the same $N_S^{\prime}=N_S$. Then the photon number per slot is measured. The slot with any click detected is the target-present slot. The errors only occur when no photon is detected among all slots. At this moment we randomly guess to make a decision. Hence the error rate is 
\begin{align}
P_E=\frac{m-1}{m}P_{m}(\bm 0,\bm 0|T)
\end{align} with the same $E,S,C$ as in the single slot case.
Similarly we have 
\be
P_{E,m}=\frac{m-1}{m}\left[\frac{1}{1+N_S(1-\sqrt{\kappa_T})}\right]^{2M}
\label{eq:PPManalyt_app}
\ee

\subsection{Numerical verification of the formula}
\label{App:verification_solvable_m}
The solutions in Eq.~(\ref{eq:analyt_app}) and (\ref{eq:PPManalyt_app}) can be summarized into a unified formula, as done in Eq.~(\ref{eq:PPManalyt}) of the main paper.

The solvable case requires the ideal channel with $\kappa_I=1, N_B=0$. 
As shown in Fig.~\ref{fig:analytical}(a), the analytical formula in Eq.~(\ref{eq:analyt_app}) fits the numerical results well in the absorption detection case and achieves the QCB. 
As shown in Fig~\ref{fig:analytical}(b), the analytical formula in Eq.~(\ref{eq:PPManalyt_app}) fits the numerical results well for the peak positioning case.

\section{Details on numerical simulation and full results}

\subsection{Monte Carlo simulation}

In the main paper, we evaluate the performance of EAAS through Monte Carlo simulations. First we randomly choose the true hypothesis $h_0$ and simulate the random outcomes $\{\bm n\}$ measured from the quantum state corresponding to $h_0$. Based on the measurement results, the conditional probabilities $p(\bm n|h)$ are generated accordingly and we make the maximum likelihood decision $\tilde{h}={\rm argmax}_h p(\bm n|h)$. Note that when there are multiple maximum hypotheses with equal probability, we make a random guess among them. Finally we figure out the frequency of $\tilde{h}\neq h_0$ as an estimation of the error probability $P_{E,m}$ of EAAS. 

The sample size, i.e., number of total simulations, determines the precision of the estimation. According to the central limit theorem, larger sample size yields lower estimation variance $\sigma_{MC}^2$ in inverse proportion. However, we have to balance the computation cost while targeting at higher precision results. As a result, a variety of sample sizes are chosen according to the specific scenarios in the main paper: Fig.~\ref{fig:contours_ideal_main}(a) with $10^{8}$ and (b)(d) with $10^6$; Figs.~\ref{fig:contours_ideal_main}(c) and ~\ref{fig:molecule} with $10^7$. With these sample sets, we have checked the validity and precision of the Monte Carlo simulation by comparing with the random guess error rate $(H-1)/H$ of the H-identical-hypothesis case in the $N_S\to0$ limit, and the analytical formulae Eqs.~(\ref{eq:analyt_app}) and (\ref{eq:PPManalyt_app}) for ideal cases $\kappa_B=1$ in Fig.~\ref{fig:analytical}. The differences all lie in the $\pm 3\sigma_{MC}$ $99.7\%$ confidence interval. In the supplemental material, we give the sample size directly in each figure involving Monte Carlo simulations.

\subsection{Optimal gain}
\label{app:opt_gain}

Indeed, in all ideal cases with no noise nor idler loss, the gain $G_0=1+N_{S0}^\prime$ nulling the signal to vacuum is given by Eq.~(\ref{eq:G0}). However the optimal gain that minimizes the error rate is not necessarily the same. We denote the optimal gain $G^{(opt)}$. Indeed, numerical results show that with most $\kappa_B,\kappa_T$ values under the ideal parameter $N_B=0, \kappa_I=1$, $G_0$ saturates QCB when $MN_S$ is relatively large, i.e. optimal gain $G^{(opt)}=G_0$. When $MN_S$ is small, $G^{(opt)}$ may deviate from $G_0$ and fluctuate in the neighbourhood.

When the idler storage efficiency $\kappa_I< 1$, although impeded from nulling the signal mode to vacuum, intuitively we expect to minimize the mean photon number of it. This requires another setting $G_{min}={\rm argmin}_G\expval{a_S^{\dagger}(G)a_S(G)}$ different from $G_0$ in Eq.~(\ref{eq:G0}).

We numerically compare the performances for a wide range of the gain, including $G_0$ and $G_{min}$, setting $M$ such that the classical optimum $P_{C,1,LB}=0.1$, we see the decay rate of quantum advantage depends on the gain $G$ shown in Fig.~\ref{fig:optimalG}. Contrary to the intuition, optimum sticks around $G_0$ in most of cases, as the performance saturates beyond it (in the presented cases $G_{min}<G_0$). In the $N_B=0$ case (Fig.~\ref{fig:optimalG} (e)(f)), when $\kappa_I<1$, the optimum can be between $G_{min}$ and $G_0$, however, the optimum error probability is only slightly smaller then that of $G_0$.

\subsection{Saturating the QCB for absorption detection case}
\label{app:more_results}
We further explore the saturation of QCB in the absorption detection case. Since QCB is only asymptotically tight, we compare the error exponents as 
\be 
R_E=-\ln(P_{E,m})/M
\ee 
and that of the QCB 
\be 
R_{QCB}=-\ln(\tilde Q_s^M/2)/M
\ee 
We also compare the classical error exponent
\be 
R_C=-\ln(P_{C,1,LB})/M.
\ee 
In Fig.~\ref{fig:quReadvsQCB-5}(a)(b), we plot the error exponent ratio $R_E/R_{QCB}$. To guarantee that we are considering the asymptotic region, we choose the mode number $M$ such that the QCB is around $10^{-5}$ in all data points. We see that the error exponent ratio $r=R_E/R_{QCB}\sim 1$ in all parameter region of $\kappa_B,\kappa_T$, verifying the optimality of the EA receiver in absence of noise ($N_B=0, \kappa_I=1$). Moreover, in Fig.~\ref{fig:quReadvsQCB-5}(c)(d) we plot the error exponent ratio $R_E/R_{C}$ in the same parameter region, where we see advantage in almost all the parameter region of $\kappa_T,\kappa_B$. Furthermore, the entangled error exponent can have a factor of $\sim8$ larger than the classical error exponent, showing a great advantage. Note that the expected monotonicity with respect to transmissivities breaks at few points, e.g. $(\kappa_T,\kappa_B)=(0.6,0)$. This is due to the discreteness of $M$ when $M$ is small [even $\sim 1$ at (0.6,0)], which results in a relative sharp change. Indeed the classical lower bounds $P_{C,1,LB}$, expected to be fixed by the proper choice on $M$, fluctuates around $0.1$ in these non-monotonic areas.

\section{Characterizing the quantum advantage in absorption detection and peak positioning}

Here we present more results related to Fig.~\ref{fig:contours_ideal_main} of the main paper.

In a practical scenario, we are interested in how much EAAS can enhance the performance, when classical schemes fail to perform well. To showcase the advantage, we fix the classical lower bound to be $0.1$ or $0.01$ and calculate the error probability achievable with EAAS. 

In Fig.~\ref{fig:contours_ideal}, we plot the error probability $P_{E,m}$ for absorption detection as a function of the transmissivities $\kappa_B$ and $\kappa_T$, while comparing it with the classical lower bounds in Eqs.~(\ref{C_Helstrom}) of the main paper. The white dashed lines divide the parameter space with/without quantum advantage. From the figure, we can see that the advantage is remarkable (several orders of magnitude for $\kappa_B,\kappa_T\simeq 1$) and also survives for a large range of parameters. In practice, when $\kappa_B\simeq 1$, we find a quantum advantage for all values of $\kappa_T$ for the problem of absorption detection.

In Fig.~\ref{fig:contours_ideal2}, we plot the error probability $P_{E,m}$ for 100-peak positioning as a function of the transmissivities $\kappa_B$ and $\kappa_T$, while comparing it with the classical lower bounds in Eq.~(\ref{LB_Gaussi}) of the main paper. From the figures, we can see that the advantage is remarkable (several orders of magnitude for $\kappa_B,\kappa_T\simeq 1$) and also survives for a relatively large range of parameters. In practice, when $\kappa_B\simeq 1$, we find a quantum advantage for $\kappa_T\gtrsim 0.2$ for 100-peak positioning. The region permitting quantum advantage is slightly smaller than the absorption detection case, which we think is due to the fact that the classical lower bounds in Eq.~(\ref{LB_Gaussi}) of the main paper is looser than Eq.~(\ref{C_Helstrom}) of the main paper.

In the analysis above, we have assumed noiseless ($N_B=0$) and lossless ($\kappa_I=1$) idler storage. In experimental practice, the presence of such noise and loss is inevitable. Therefore we also study how the quantum advantage varies with idler loss $\sqrt{1-\kappa_I}$ and noise $N_B$ in Figs.~\ref{fig:kINbcontour} and~\ref{fig:kINbcontour2} for $N_S=0.1$ and $N_S=1$. Again we fix the classical lower bound and plot the error probability of EAAS for $\kappa_T=0.75$ and $\kappa_B=0.95$. Regions on the left-down side of the dashed lines show quantum advantage. We see that the advantage is robust against idler loss and channel noise.

\section{Details on general spectrum recognition}

Here we give the transmissivities data that are used in our calculations of wine-tasting and drug-testing.

\begin{figure*}
    \centering
    \subfigure[Error rate in base 10 logarithmic scale.]{
    \includegraphics[width=0.45\textwidth]{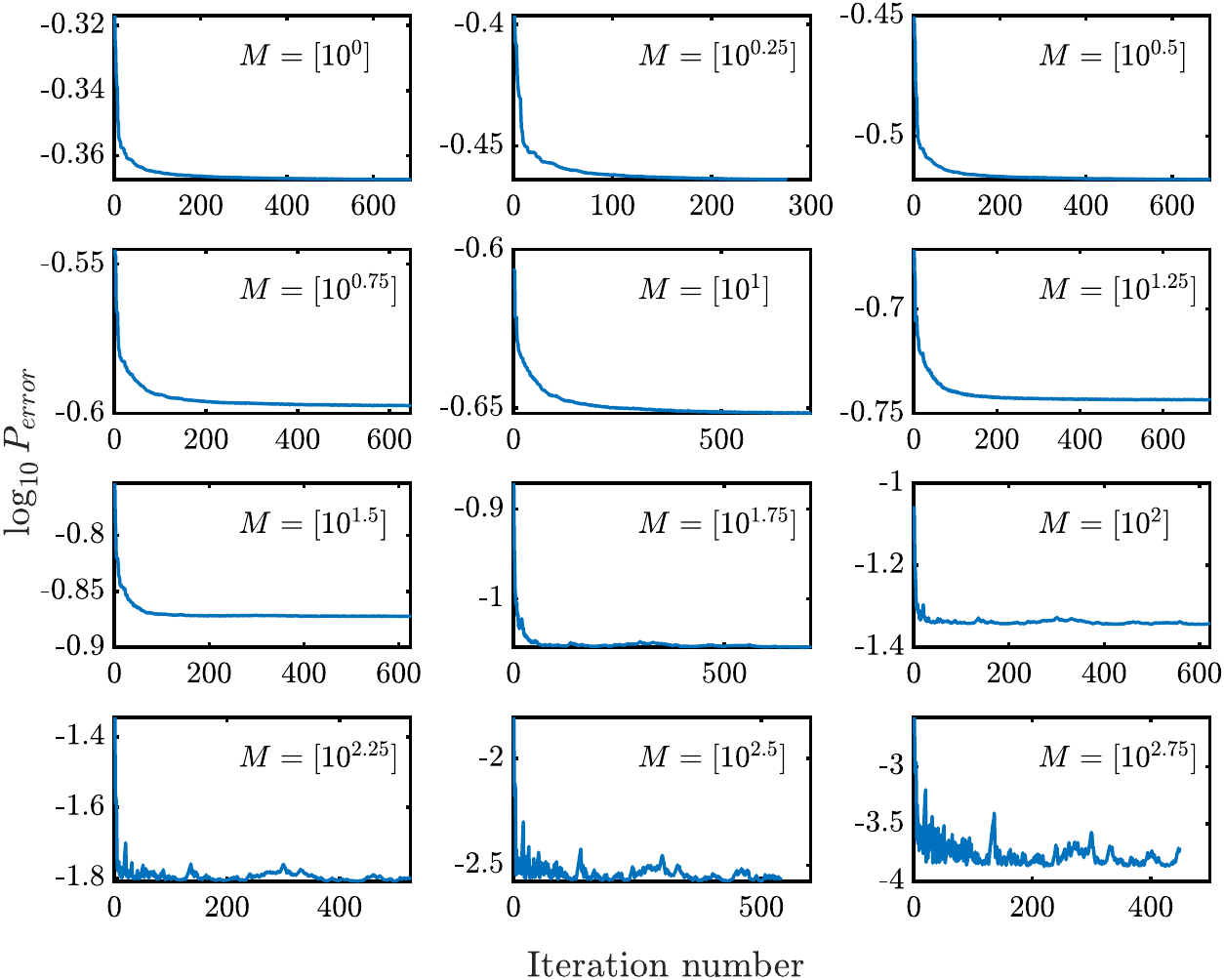}
    }
    \subfigure[Energy distribution $\bm N_S$.]{
    \includegraphics[width=0.45\textwidth]{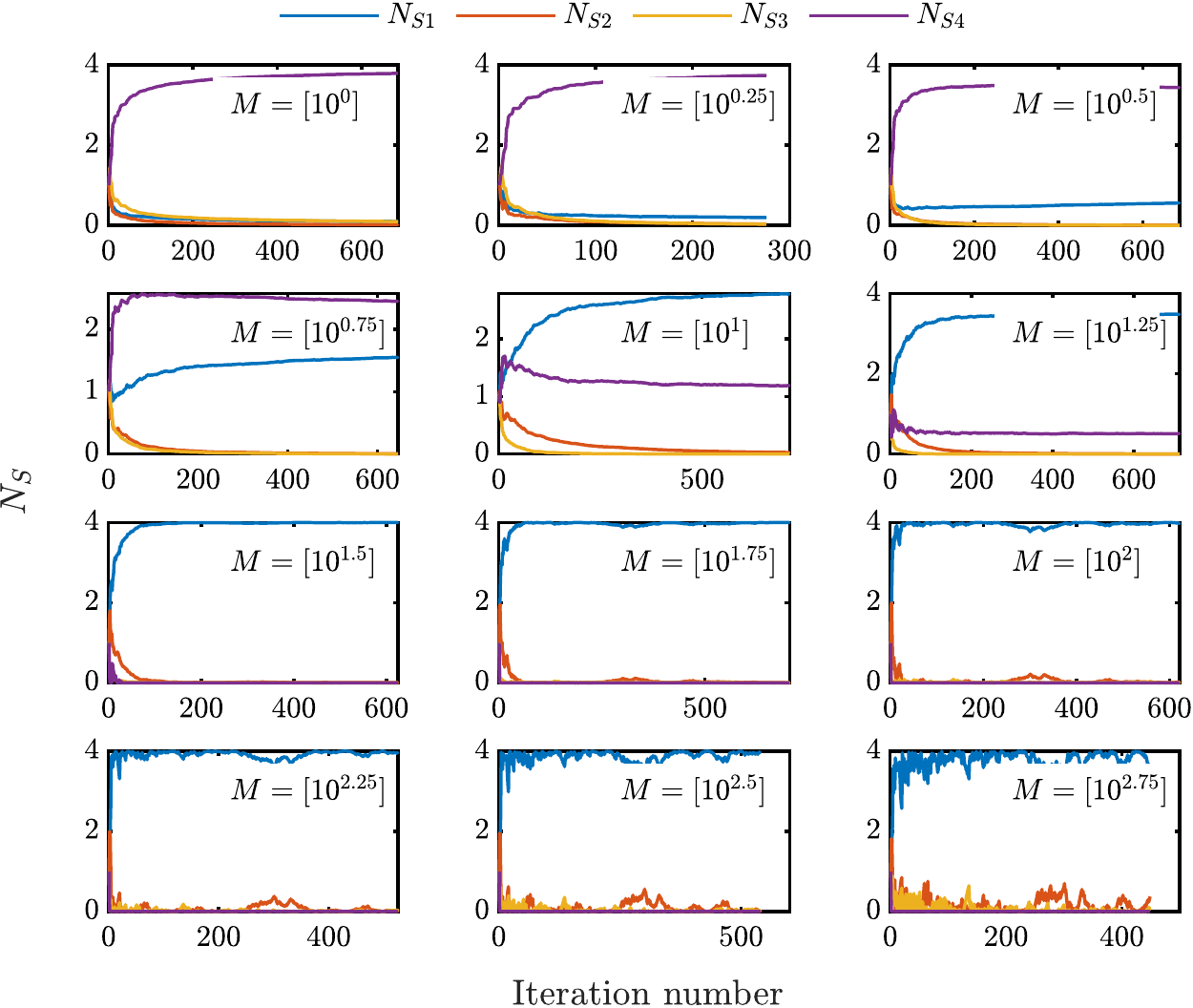}
    }
    \caption{Logarithmic scaled learning curve of the error rate optimization over the source energy distribution in the noiseless wine tasting scenario. Average mean photon number per mode $\sum_{\ell=1}^4 N_{S\ell}/4=1$, noise $N_B=0$.
    We find the energy concentrates on the first frequency slot when $M$ is small, and the fourth frequency slot when $M$ goes sufficiently large. It verifies the intuition that suggests the energy be concentrated on the slot with the largest error exponent which dominates the decay of error rate.
    \label{fig:alcoh_optNs}}
    \subfigure[Error rate in base 10 logarithmic scale.]{
    \centering
    \includegraphics[width=0.45\textwidth]{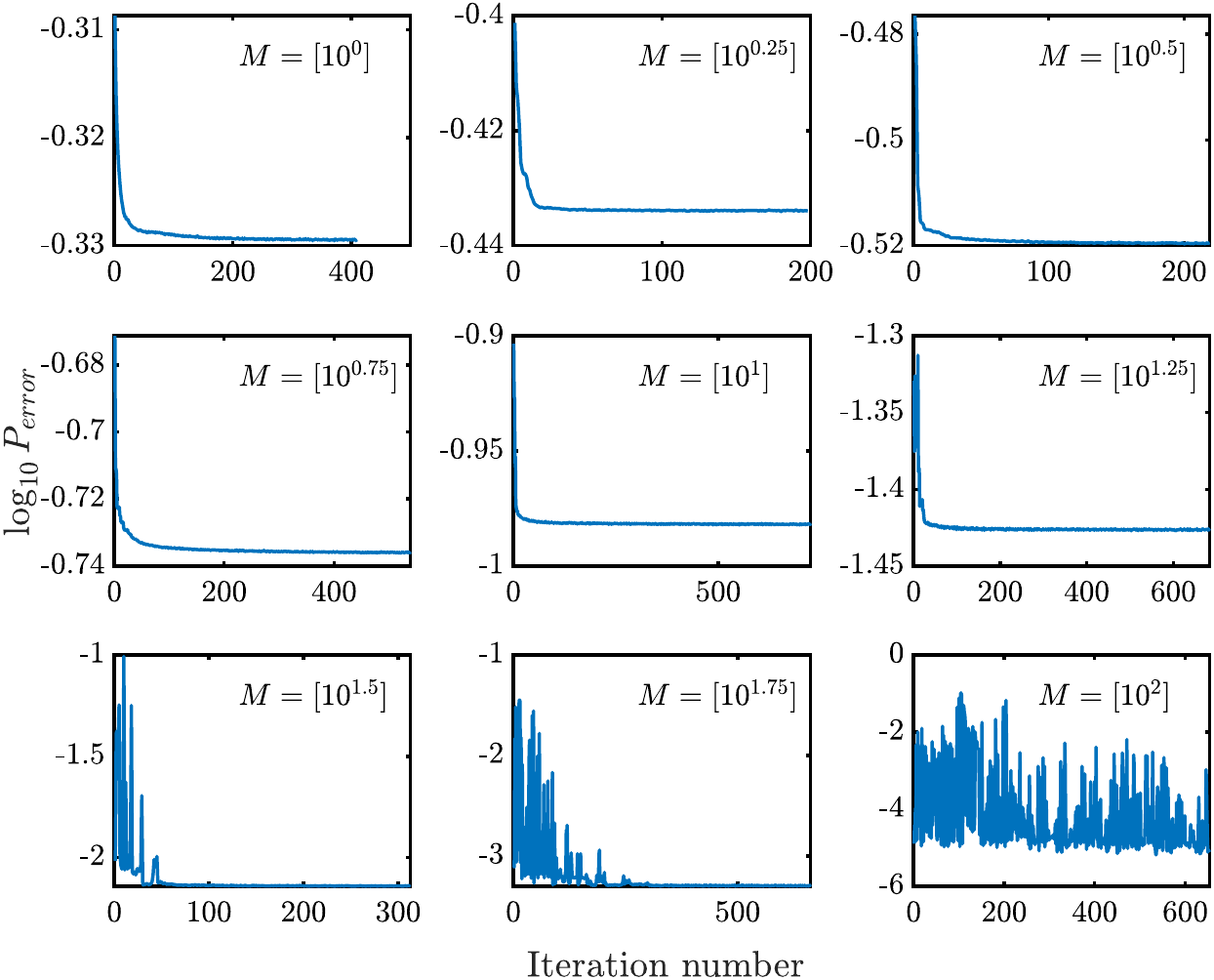}
    }
    \subfigure[Energy distribution $\bm N_S$.]{
    \centering
    \includegraphics[width=0.45\textwidth]{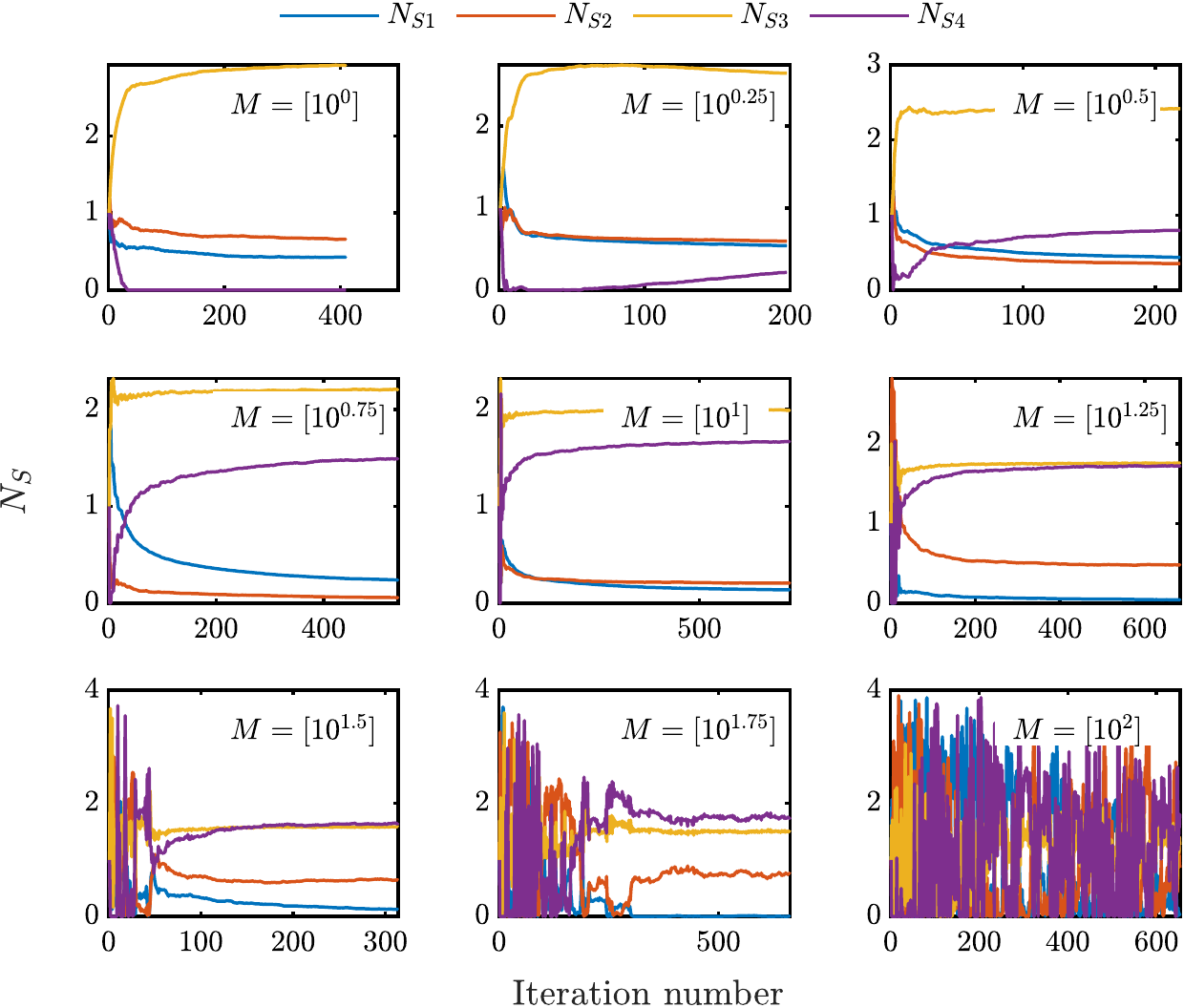}
    }
    \caption{Learning curve of the error rate optimization over the source energy distribution in the noiseless drug testing scenario. Average mean photon number per mode $\sum_{\ell=1}^4 N_{S\ell}/4=1$, noise $N_B=0$. The large fluctuation in the last two subfigures of $M=[10^{1.75}]$ and $M=[10^2]$ is due to the Monte Carlo simulation error. The concentration of energy on a single frequency slot predicted by the classical bound is absent here, rather the optimum energy distribution is non-trivial on all slots. 
    \label{fig:drug_optNs}}
\end{figure*} 
\begin{figure*}
    \subfigure[Error rate.]{
    \centering
    \includegraphics[width=0.45\textwidth]{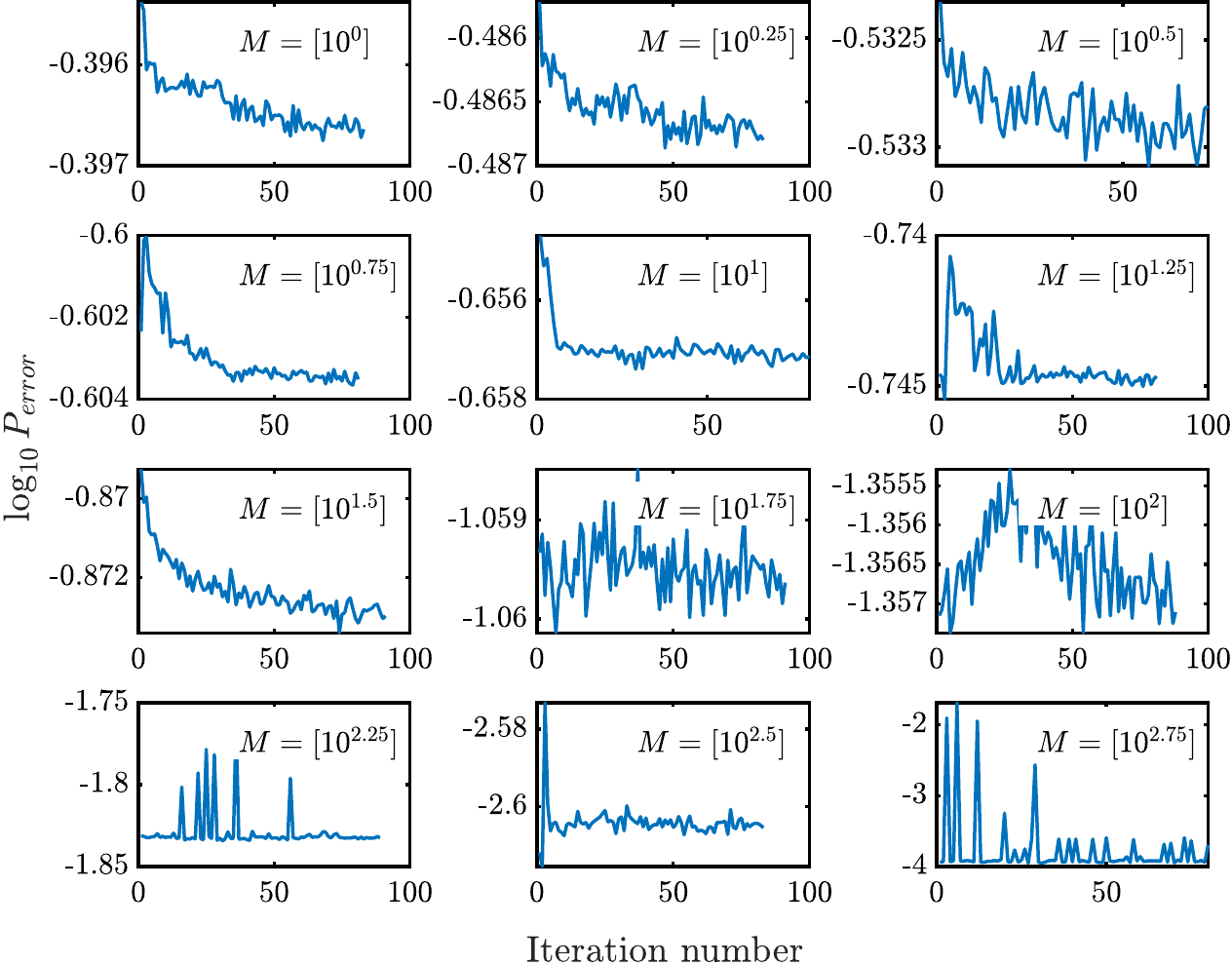}
    }
    \subfigure[Gain distribution $\bm G$]{
    \centering
    \includegraphics[width=0.45\textwidth]{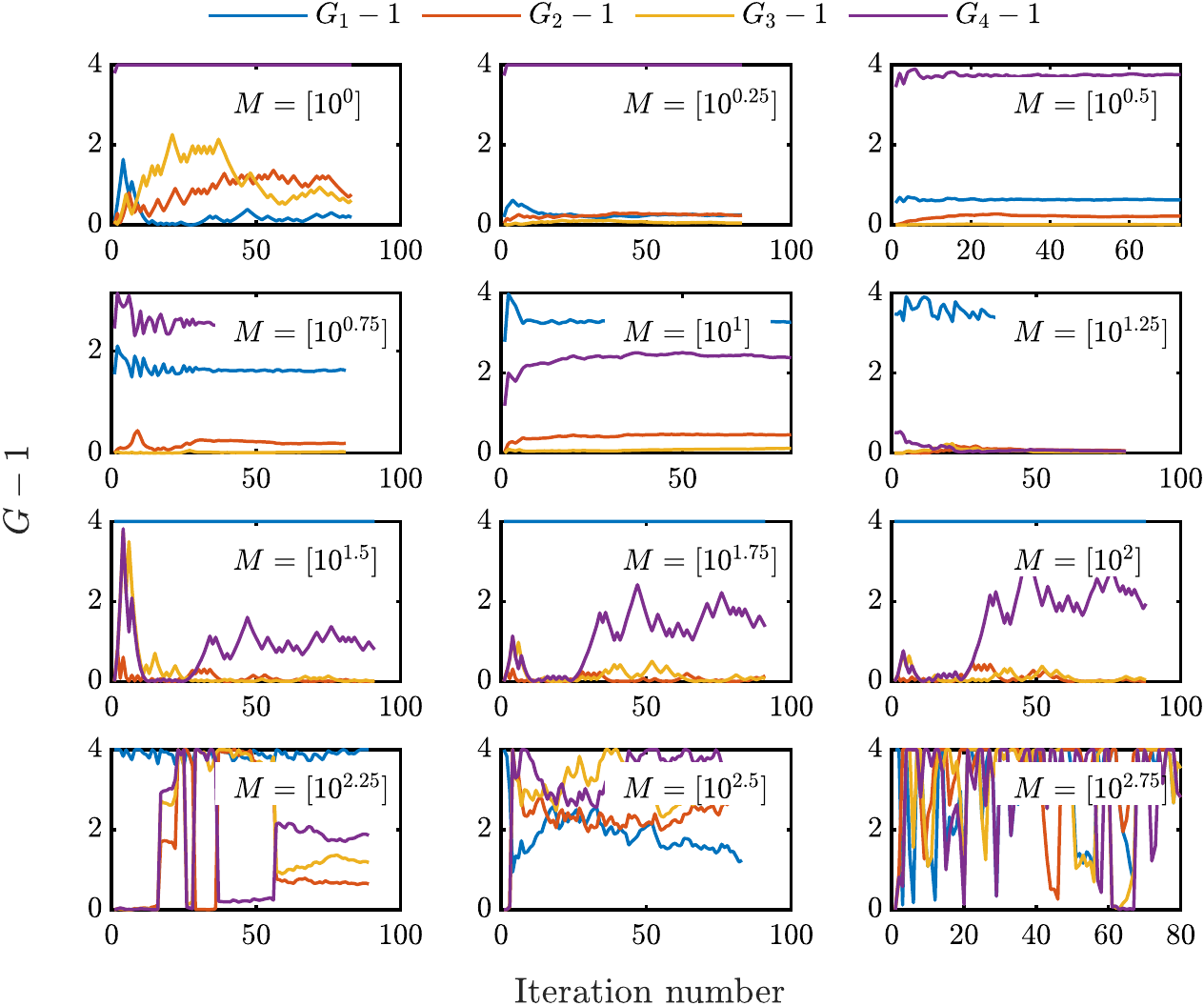}    
    }
    \caption{Learning curve of the error rate optimization over the receiver gain in the noiseless wine tasting scenario. Average mean photon number per mode $\sum_{\ell=1}^4 N_{S\ell}/4=1$, noise $N_B=0$. The gains are restricted by $\bm G-1\leq 4$. \label{fig:alcoh_optG}}
    \subfigure[Error rate.]{
    \centering
    \includegraphics[width=0.45\textwidth]{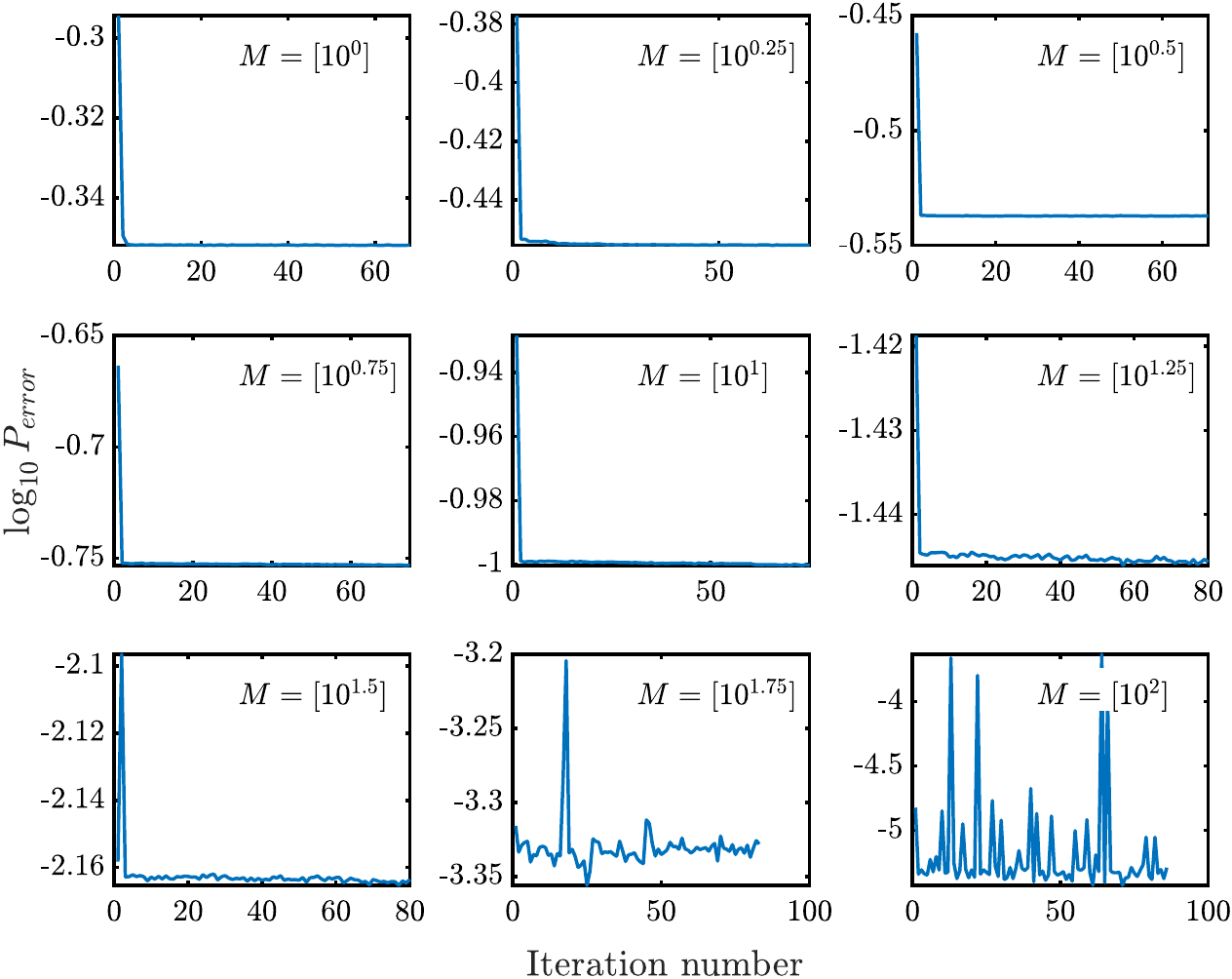}
    }
    \subfigure[Gain distribution $\bm G$]{
    \centering
    \includegraphics[width=0.45\textwidth]{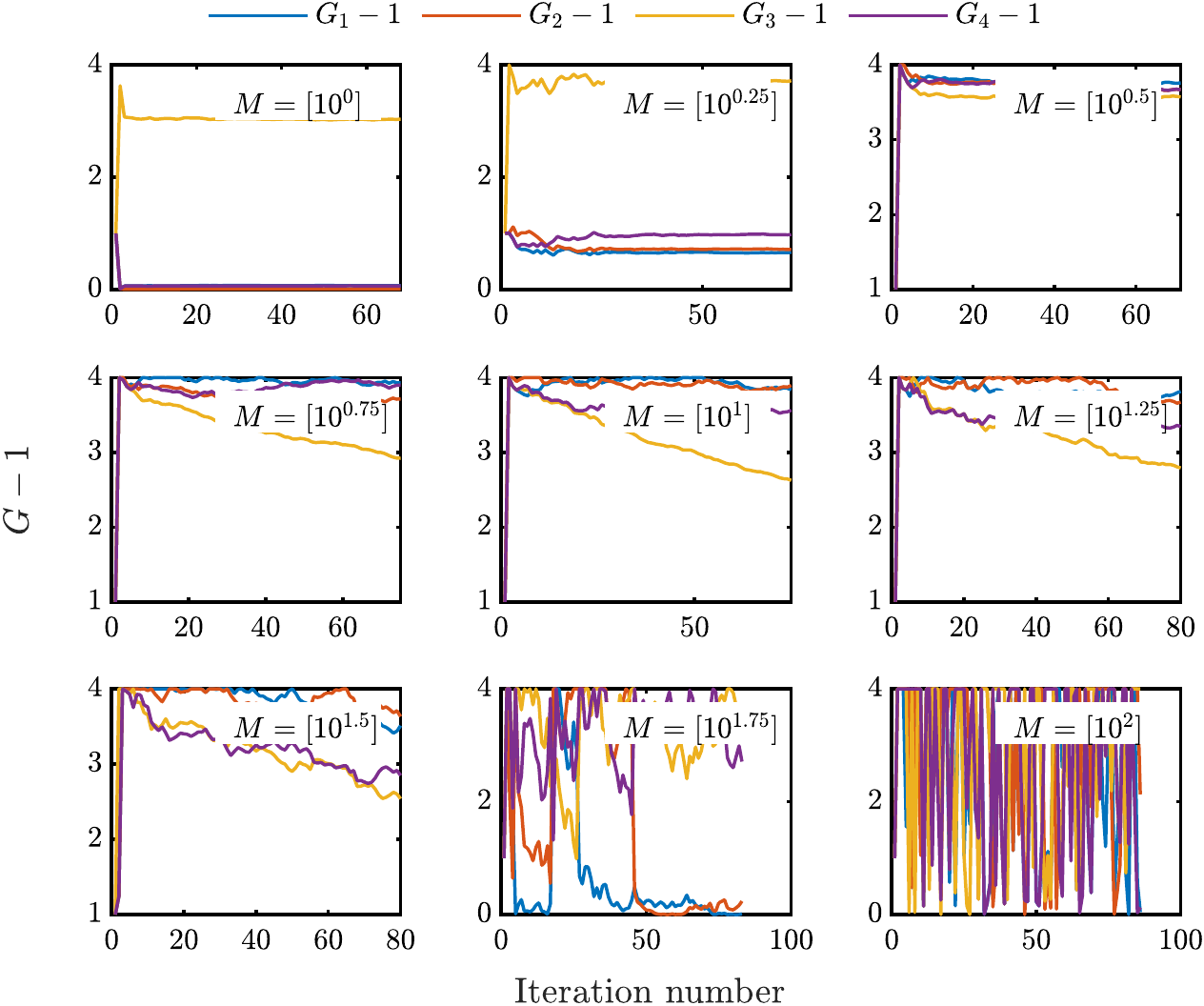}   
    }
    \caption{Learning curve of the receiver gain optimizing the receiver gain in the noiseless drug testing scenario. Average mean photon number per mode $\sum_{\ell=1}^4 N_{S\ell}/4=1$, noise $N_B=0$. The gains are restricted by $\bm G-1\leq 4$. 
    \label{fig:drug_optG}}
\end{figure*}

For the wine tasting, there are $H=3$ molecules---methanol ($h=1$), ethanol ($h=2$) and ethanal ($h=3$). To discretize the spectra so that we can perform numerical simulation, samples are tested at four frequency slots 500, 1050, 1400, 1800 ${\rm cm}^{-1}$ (wavelength 20, 9.5, 7.1, 5.6$\mu m$), with the transmissivities averaged in the span of $\pm 100\,{\rm cm}^{-1}$. The discrete spectrum can be described by
$\kappa^{(h)}_\ell=K_{\ell h}$, where the overall data can be represented as an $m\times H$ matrix
\be
K=
\begin{pmatrix}
   0.9460  &  0.9749  &  0.7853\\
    0.5659 &   0.6218 &  0.6846\\
    0.7503 &   0.7622 &   0.4683\\
    0.9737 &   0.9891 &   0.4165\\
\end{pmatrix}.
\label{K_matrix_wine}
\ee
Here each column gives the transmissivities of ${\bm \kappa}^{(h)}$ for a fixed hypothesis $h=1,2,3$. The four rows in each column corresponds to the $m=4$ spectrum slots.

For the drug-testing case, the molecules are phenyl salicylate ($h=1$), methyl salicylate ($h=2$), and benzoic acid ($h=3$). Similarly, samples are tested at four frequency slots 500, 100, 1500, 2000 ${\rm cm}^{-1}$ (wavelength 20, 10, 6.67, 5$\mu m$), with the transmissivities averaged in the span of $\pm 100\,{\rm cm}^{-1}$. We have
$\kappa^{(h)}_\ell=K_{\ell h}$, with
\be
K=
\begin{pmatrix}
 0.9613    &0.9002  &  0.8093\\
        0.9215  &  0.8749  &  0.7427\\
        0.8360  &  0.4002  &  0.7556\\
        0.9867  &  0.8749  &  0.8522
\end{pmatrix}.
\label{K_matrix_drug}
\ee

\begin{figure*}
    \centering
    \subfigure[Wine tasting.]{
    \centering
    \includegraphics[width=0.25\textwidth]{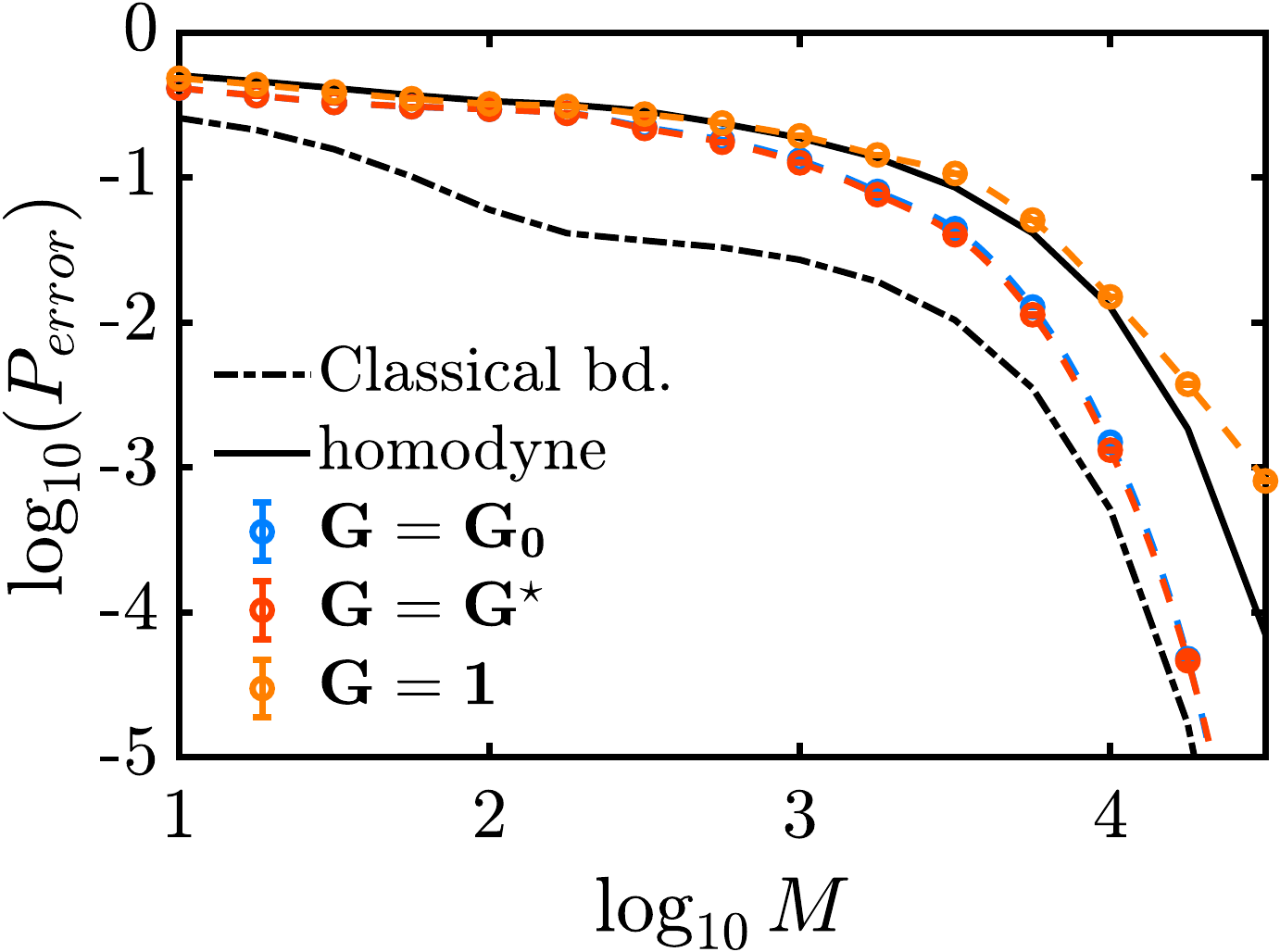}
    }
    \subfigure[Drug testing.]{
    \centering
    \includegraphics[width=0.25\textwidth]{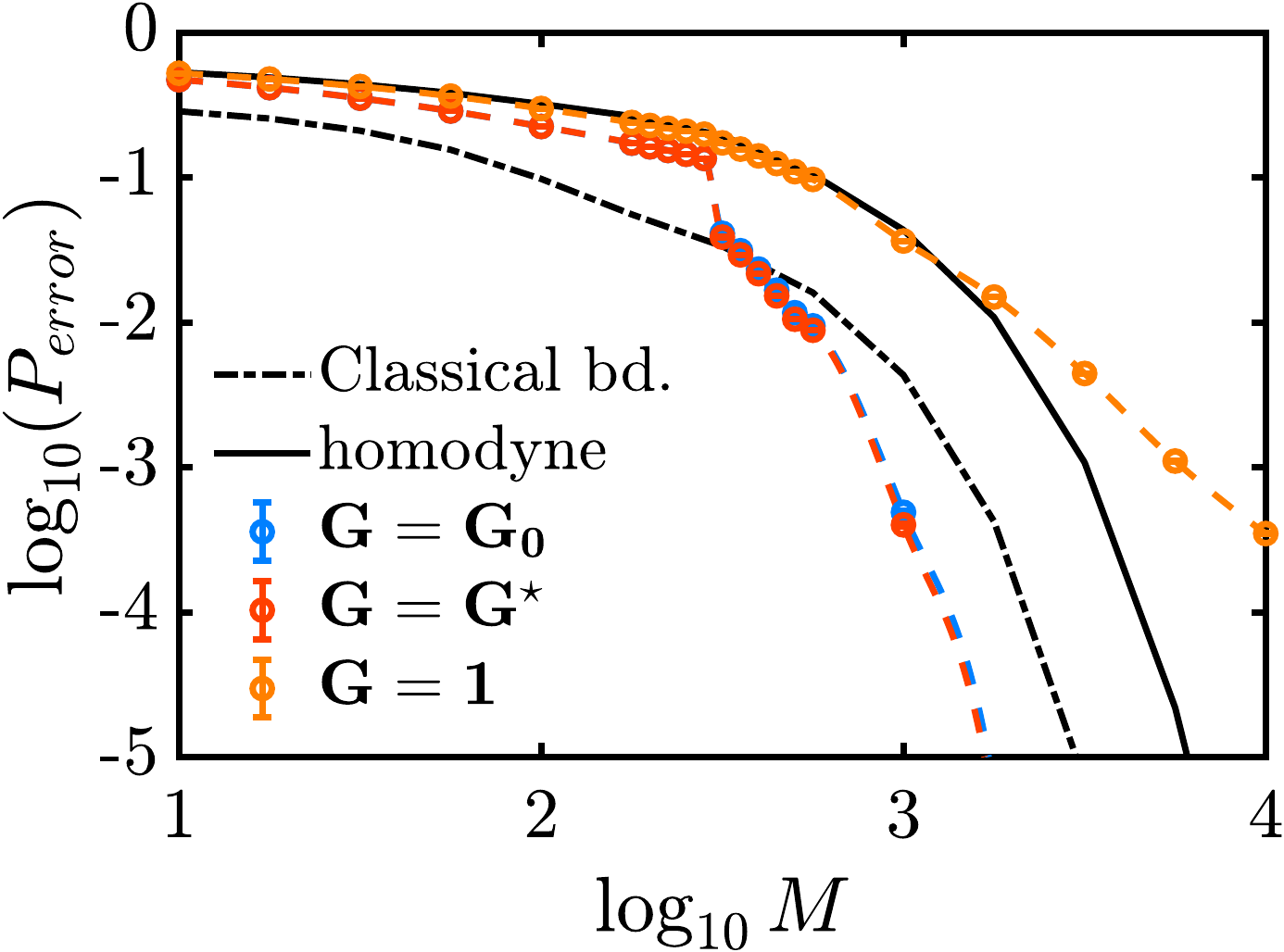}   
    }
    \subfigure[Optimized energy distribution of the drug testing.]{
    \centering
    \includegraphics[width=0.25\textwidth]{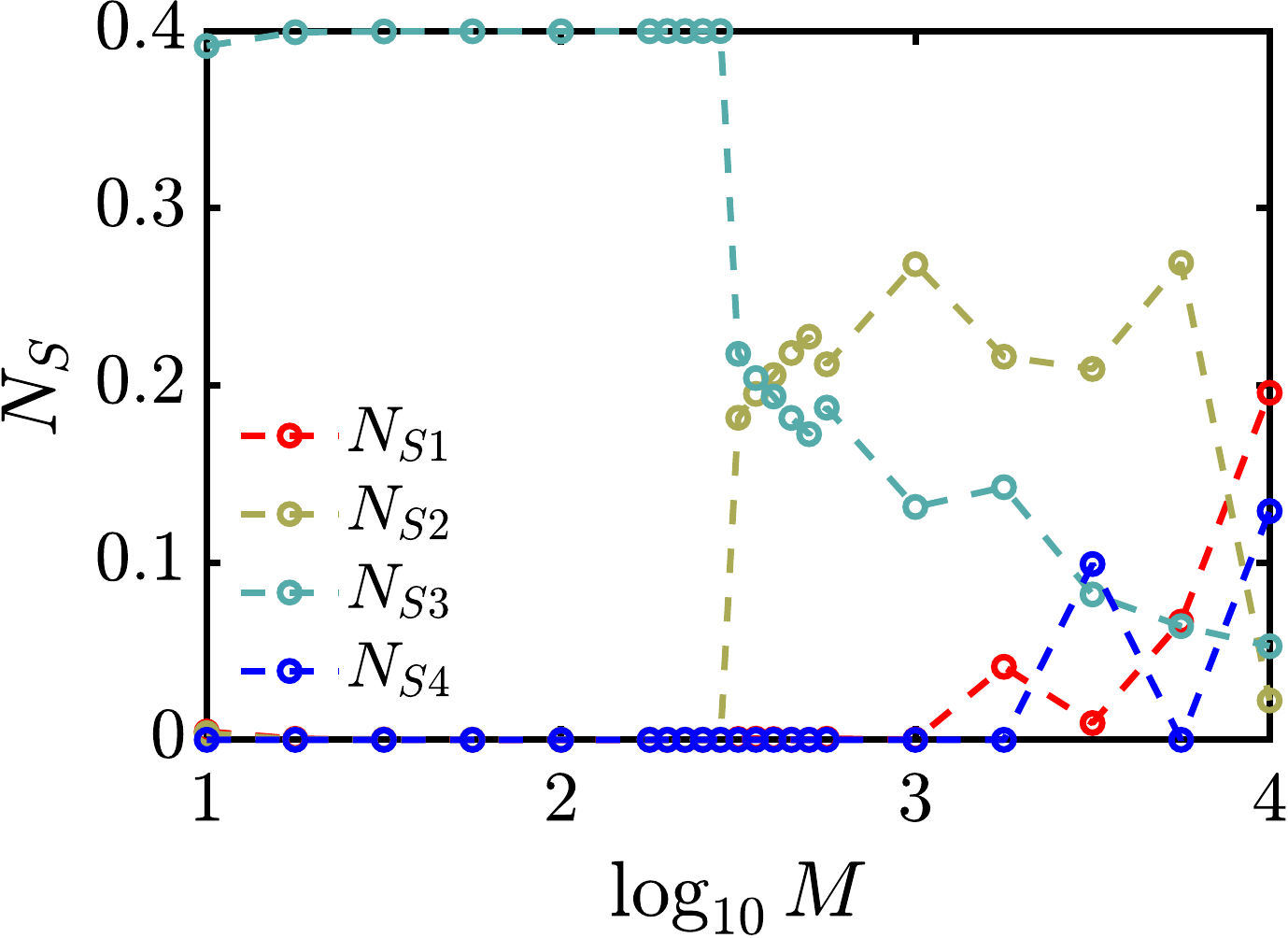}   
    }
    \caption{Error rate of the receiver in the noisy scenario with thermal photon number $N_B=0.1$. The energy distribution of the source is optimized under the constraint on the average mean photon number per mode $\sum_{\ell=1}^4 N_{S\ell}/4=1$. We compare energy-optimized EAAS under three gain settings: simply adopting Eq.~\eqref{eq:G0} in each frequency slot to obtain $\bm G=\bm G_0=1+\bm N_{S0}^\prime$ (dashed blue), numerically optimized gain $\bm G=\bm G^\star$ (dashed red), OPA-absent case $\bm G=1$ (dashed orange), along with the classical lower bound Eq.~\eqref{LB_numerical_app} (dot-dashed black) and homodyne detection (solid black). In the numeric optimization of $\bm G^\star$ the gains are restricted by $\bm G-1\leq 4$. Subplot (c) shows the optimized energy distribution which is prescribed in the optimization in subplot (b), where the leap at $M\sim 10^{2.5}$ accounts for the same leap in the performance in (b).
    \label{fig:gainopt_noisy}}
\end{figure*}

We evaluate the performance of our receiver through numerical simulations, with optimization over energy and gain. We generate the measurement results through Eq.~\eqref{condtional_prob_app}, and perform maximum likelihood decision. And the error rates are obtained by calculating the frequencies of error happening.

Noticing that the general patterns in Eq.~\eqref{K_matrix_wine} and~\eqref{K_matrix_drug} break the symmetry between different (frequency) slots, symmetric strategies with uniform parameters over all slots are unlikely to be the optimum; therefore we expect optimizing the parameters, including the source mean photon numbers $\bm N_S=[N_{S1},N_{S2},N_{S3},N_{S4}]^T$ and the receiver gains $\bm G=[G_1,G_2,G_3,G_4]^T$ over the four frequency slots, to achieve further quantum advantages. Considering the large dimension of parameters in the future application with more frequency slots, we apply the constrained simultaneous perturbation stochastic approximation (SPSA) algorithm \cite{spall1988stochastic,sadegh1997constrained} to accelerate the optimization. As a variation of stochastic gradient descent methods, SPSA does not guarantee to always find the global optima. First, we will perform optimization over the energy distribution, and then we will further optimize over the gain. We find that in general energy optimization leads to improved quantum advantages, gain optimizations offers slight improvement in the noiseless case and much better improvement in the noisy case.

\subsection{Energy distribution of the source}
In the noiseless scenario, Figs.~\ref{fig:alcoh_optNs} and~\ref{fig:drug_optNs} show the learning curve of the optimization over the source energy distribution, with the average mean photon number per mode fixed to $\sum_{\ell=1}^4 N_{S\ell}/4=1$. From the subplots (b) in both figures, we see that the optimal energy distributions are far from uniform as expected. We achieve an appreciable one-order-of-magnitude improvement when $M$ is large, which guarantees the advantage over the classical lower bound in Fig.~\ref{fig:molecule} of the main text. 

As to the energy distribution, the classical lower bound in Eq.~\eqref{LB_general} of the main paper gives a pretty good intuition for the optimization, although the lower bound itself is likely to be loose. Specifically, the classical bound consists of a few exponential terms with respect to the difference between the transmitted energys $(\sqrt{\kappa_i}-\sqrt{\kappa_j})^2MN_S$ through two channels $i,\,j$, and the term with the largest error exponent dominates the performance asymptotically when $M$ is sufficiently large. Hence a fair guess of the optimal strategy is to allocate most of the energy to the slot associated with the dominant term. However, in Fig.~\ref{fig:alcoh_optNs}(b), we see that the energy tends to concentrate on the first slot when M increases, which is different from the classical optimum (the second slot). Note that the transmissivities of the first slot are relatively higher and closer to unity, this divergence from the classical prediction is likely due to the gap between the quantum scenario here and the conventional classical region, viz. that the entanglement-assisted advantage over the classical optimum surges as the transmissivity increases close to unity. This can also be seen in Fig.~\ref{fig:drug_optNs}(b), the energy was scattered on the four modes without any dominant slot, instead of being concentrated on the second slot as predicted by the classical optimum. The illustrated differences from the intuition based on classical schemes demonstrate the novelty of the entanglement-assisted scenario again.

\subsection{Gain distribution of the receiver}
With the same noiseless environment setting, we optimize the gain distribution on top of the optimal energy distribution achieved above. Figs.~\ref{fig:alcoh_optG} and~\ref{fig:drug_optG} give the learning curve of the optimization on the receiver gain distribution. The gains are restricted per slot by $\bm G-1\leq 4 N_S=4$, as the gain of OPA is limited by the nonlinearity of the crystal in practice. Compared with the optimization on energy distribution, we see a slight improvement by further optimizing the gains in this scenario.

The optimization of the gain distribution is more challenging than that of the energy distribution. As shown in Fig.~\ref{fig:optimalG} (e-h) for the binary hypothesis-testing case, the gradient of error rate with respect to the gain is close to zero almost everywhere. A large fluctuation is present in the convergence process due to the small gradient-to-noise ratio.

By contrast, gain optimization is likely to achieve a significant improvement in the noisy scenario. To demonstrate the influence of noise, we include a uniform thermal background with mean photon number $N_B=0.1$ into the channel. Following the same procedure, we first optimize the energy distribution for the noisy case. Here the presence of noise leads to a sharp increase in the computational complexity of the photon statistics. Consequentially, we compromise the average mean photon number per mode down to $\sum_{\ell=1}^4 N_{S\ell}/4=0.1$.  Meanwhile, we correspondingly scale the range of copy number $M$ of interest by a factor 10. Fig.~\ref{fig:gainopt_noisy} illustrates the gap between the different gain settings. In both the wine-tasting and drug-testing cases, we see a substantial advantage of the gain optimization over the OPA-absent case as expected. Interestingly, a trivial choice of $\bm G=\bm G_0=1+\bm N_{S0}^\prime$ according to Eq.\eqref{eq:G0} is sufficient to yield nearly the same improvement. This provides a neat rule-of-thumb estimation of the optimum gain in the practical implementation.


%

\end{document}